\documentclass[aps,prd,twocolumn,superscriptaddress]{revtex4-1}
\usepackage[utf8]{inputenc}
\usepackage{amsmath}
\usepackage{amsfonts}
\usepackage{hyperref}
\usepackage{cleveref}
\usepackage{amssymb}
\usepackage{graphicx}
\usepackage{orcidlink}
\usepackage{mathrsfs}
\usepackage{tikz}
\usetikzlibrary{decorations.markings}
\usepackage{gnuplot-lua-tikz}
\usepackage{color}
\newcommand{\dtk}{\textit{d}\tilde k}

\newcommand{\rwb}{\sqrt{m\rv}}

\newcommand{\sllm}{\textit{S}_{l,l-1}}
\newcommand{\slls}{\textit{S}_{l,l-1}^{\mathrm{Sty}}}
\newcommand{\sllt}{\textit{S}_{l,l-1}^{\mathrm{Tra}}}
\newcommand{\sllf}{\textit{S}_{l,l-1}^{\mathrm{Ftp}}}
\newcommand{\sllb}[2]{\textit{S}_{l,l-1}^{#1, #2}}
\newcommand{\bessJ}[2]{\textit{j}_{#1}(#2)}
\newcommand{\bessY}[2]{\textit{y}_{#1}(#2)}
\newcommand{\bh}[3]{\textit{h}_{#2}^{(#1)}(#3)}
\newcommand{\hone}[2]{\textit{h}_{#1}^{(1)}(#2)}
\newcommand{\htwo}[2]{\textit{h}_{#1}^{(2)}(#2)}
\newcommand{\res}[2]{\text{Res}\left(#1 , #2\right)}

\newcommand{\green}[2]{\frac{k\textit{h}_{l}^{(#1)}( \tk) \textit{h}_{l-1}^{(#2)}(\tk)}{k^4/\rv^2-m^2} }
\newcommand{\resResult}[3]{\frac{\rv}{4 m}\,\textit{h}_{l}^{(#1)}(#3)\,\textit{h}_{l-1}^{(#2)}(#3) }
\def\vk{\mathbf{k}}
\def\vr{\mathbf{r}}
\def\vu{\mathbf{u}}
\def\vv{\mathbf{v}}
\def\tvk{\tilde{\mathbf{k}}}
\def\tvr{\tilde{\mathbf{r}}}
\def\tvu{\tilde{\mathbf{u}}}

\def\tk{\tilde k}
\def\tkmin{\tilde k_{\mathrm{min}}}
\def\ttau{\tilde\tau}
\def\tomega{\tilde\omega}
\def\vvl{\mathbf{L}}

\def\vvf{\mathbf{F}}
\def\hvr{\hat{\mathbf{r}}}
\def\hvz{\hat{\mathbf{z}}}
\def\hpv{\hat{\boldsymbol{\varphi}}}
\def\fdf{\vvf_\text{DF}}
\newcommand{\intk}{\int_{\vk}}
\newcommand{\intw}{\int_\omega}
\newcommand{\inttk}{\int_{\tvk}}
\newcommand{\inttw}{\int_{\tomega}}
\newcommand{\ls}{\lambdabar_\sigma}
\newcommand{\lv}{\lambdabar_\Omega}
\newcommand{\lQ}{\lambdabar_Q}
\newcommand{\rs}{R_\sigma}
\newcommand{\rv}{R_\Omega}
\newcommand{\rQ}{R_Q}
\newcommand{\ms}{\mathcal{M}_\sigma}
\newcommand{\mQ}{\mathcal{M}_Q}
\newcommand{\mg}{\mathcal{M}_g}
\newcommand{\am}{\alpha^{\mathrm{M}}}
\newcommand{\als}{\alpha^{\mathrm{LS}}}
\newcommand{\vrcm}{\vr_\text{CM}}
\newcommand{\agw}{a_\text{GW}}
\def\pc{\ {\rm pc}}
\def\Kpc{\ {\rm Kpc}}
\def\eV{\ {\rm eV}}
\def\yr{\ {\rm yr}}
\def\msun{\ M_\odot}
\def\mpppc{\ M_\odot\, {\rm pc}^{-3}}
\def\kms{\ {\rm km\,s^{-1}}}

\def\actaa{\ref@jnl{Acta Astron.}}      

\begin{document}

\title{Dynamical Friction in fuzzy dark matter: circular orbits}

\author{Robin Buehler\, \orcidlink{0000-0003-3073-9036}}
\email{robinbuehler@campus.technion.ac.il}
\affiliation{Physics department, Technion, Haifa 3200003, Israel}

\author{Vincent Desjacques\,\orcidlink{0000-0003-2062-8172}}
\email{dvince@physics.technion.ac.il}
\affiliation{Physics department, Technion, Haifa 3200003, Israel}

\begin{abstract}
    Dynamical frictional (DF) is the gravitational force experienced by a body moving in a medium as a result of its density wake. In this work, we investigate the DF acting on circularly-moving perturbers in fuzzy dark matter (FDM) backgrounds. After condensation in the early Universe, FDM is described by a single wave function satisfying a Schr\"odinger-Poisson equation. An equivalent, hydrodynamic formulation can be obtained through the Madelung transform. Here, we consider both descriptions and restrict our analysis to linear response theory. We take advantage of the hydrodynamic formulation to derive a fully analytic solution to the DF in steady-state and for a finite time perturbation (corresponding to a perturber turned on at $t=0$). We compare our prediction to a numerical implementation of the wave approach that includes a non-vanishing FDM velocity dispersion $\sigma$. Our 
    solution is valid for both a single and a binary perturber in circular motion as long as $\sigma$ does not significantly exceed the orbital speed $v_\text{circ}$. While the short-distance Coulomb divergence of the (supersonic) gaseous DF is no longer present, DF in the FDM case exhibits an infrared divergence which stems from the (also) diffusive nature of the Schr\"odinger equation. Our analysis of the finite time perturbation case reveals that the density wake produced by perturbers diffuses through the FDM medium until it reaches its outer boundary. Once this transient diffusive regime is over, both the radial and tangential DF oscillate about the steady-state solution with a decaying envelope. Steady-state is never achieved. We discuss two astrophysical applications of our results: we revisit the DF decay timescales of the 5 Fornax globular clusters, and point out that the inspiral of compact binary may stall because the DF torque about the binary center-of-mass sometimes flips sign to become a thrust rather than a drag.
\end{abstract}

\maketitle

\section{Introduction}
\label{sec:introduction}

The lack of evidence for weakly interactive massive particles at current collider experiments strongly advocates the exploration of alternative dark matter scenarios. In the fuzzy dark matter (FDM) scenario \citep{Baldeschi:1983mq,Sin:1992bg,hu/etal:2000}, dark matter is in the form of ultra-light bosons that are an extrapolation of the QCD axion 
\citep{peccei/quinn:1977,wilczek:1978,weinberg:1978}
down to very small masses \cite[see][for a review]{marsh:2016}.
While QCD axions have masses in the range $10^{-10} \lesssim m_a \lesssim 10^{-3}$ eV, the FDM particles can have masses as low as $m_a\gtrsim 10^{-21}$ eV without spoiling cosmic microwave background and large scale structure constraints \cite{irsic/etal:2017,armengaud/etal:2017,hlozek/etal:2017}. However, recent astrophysical constraints from a variety of systems and scales suggest a lower limit $m_a\gtrsim 10^{-19} - 10^{-18}$ eV \cite{marsh/pop:2015,calabrese/spergel:2016,gonzalez/etal:2017,marsh/niemeyer:2018,desjacques/nusser:2019,bar/etal:2019} (see, however, \cite{demartino/etal:2020}). Higher axion masses can also be probed with the super-radiant instability of spinning black holes \cite{cardoso/etal:2018}.

Light bosons generically undergo Bose-Einstein condensation in the early Universe \cite{boehmer/harko:2007,sikivie/yang:2009,guth/etal:2015}, after which their spatial distribution is characterized by a (classical) wave function satisfying a nonlinear Schr\"odinger equation. This condensate behaves like non-relativistic, cold dark matter (CDM) on scales larger than the de Broglie wavelength of the particles \cite[e.g.][]{hwang/noh:2009}. Therefore, gravitationally bound structures form hierarchically like in CDM cosmologies, though virialized FDM halos grow solitonic cores at their center \cite{ruffini/bonazzola:1968,chavanis:2011,schive/etal:2014a,schwabe/etal:2016,mocz/etal:2017,veltmaat/etal:2018}. These dense central cores are surrounded by a large atmosphere of fluctuating granules \citep{schive/etal:2014b,hui/etal:2017,Bar2019}.

The motion of extended or compact objects in a FDM background generates a dynamical friction (DF) as in any other ambient medium \cite{chandrasekhar:1943,ostriker:1999}. The gravitational field of a perturber moving in a discrete or continuous medium induces a density fluctuation or wake. DF is the gravitational force exerted on the perturber by the density wake.
The studies of \cite{chandrasekhar:1943,tremaine/weinberg:1984} considered a perturber linearly moving in a collisionless medium, whereas \cite{dokuchaev:1964,ruderman/spiegel:1971,rephaeli/salpeter:1980,just/kegel:1990,ostriker:1999,sanchez/brandenburg:2001,kim/etal:2007,lee/stahler:2011,vicente/etal:2019,DesNuBu21,macleod/etal:2022} focused on a gaseous medium. Dynamical friction in FDM and Bose-Einstein condensates has been explored only recently in, e.g. \cite{hui/etal:2017,Bar2019,berezhiani/eal:2019,Lan2020,annulli/etal:2020,traykova/clough/etal:2021,rodrigo/cardoso:2022}. In particular, \cite{hartman/etal:2021} considered DF in a Bose-Einstein condensate (BEC) with weak self-interactions, while \cite{annulli/etal:2020,wang/easther:2022} numerically investigated the DF induced by linearly and circularly moving point masses taking into account the self-gravity of the FDM background.

In this work, we will investigate the DF acting on point masses in circular motion using the analytical approach outlined in \cite{DesNuBu21}. While the validity of this approach is restricted to linear response theory, it provides a versatile tool to explore DF across a wide parameter range. Here, we apply this methodology to point mass perturbers circularly moving in FDM backgrounds. Starting from the hydrodynamic (Madelung) formulation of the Gross-Pitaievskii-Poisson (GPP) system, we will derive new analytical predictions which we compare to the DF computed in the wave (Lippmann-Schwinger) approach. We will also consider finite time perturbations (i.e. the perturber is turned on at $t=0$) and explore the convergence to steady-state. 

Our paper is organized as follows. Sec.~\S\ref{sec:scales} introduces the key scales and dimensionless parameters that control DF for the systems of interest. Sec.~\S\ref{sec:GPP} summarizes the application of linear response theory to a FDM background, and presents numerical solutions for density wakes. 
In Sec.~\S\ref{sec:DF}, we briefly recapitulate the approach of \cite{DesNuBu21} before we spell out the derivation of DF for circularly-moving perturbers in FDM backgrounds and perform a number of numerical tests.
In Sec.~\S\ref{sec:discussion}, we discuss some astrophysical implications of our results pertaining, in particular, to the infall times of globular clusters and the stagnation of binary inspirals. We conclude in Sec.~\S\ref{sec:conclusion}.
Technical details of the derivation of our analytic results are summarized in a few appendices.

\section{Characteristic scales}
\label{sec:scales}

For convenience, we begin by introducing our notations and the key dimensionless parameters used throughout the paper. We follow somewhat the convention of \cite{Lan2020} to facilitate the comparison with their results.

The point-like perturber of mass $M$ moves on circular orbits of radius $r_0$ and frequency $\Omega$. These physical quantities will be used to define dimensionless variables labeled with a tilde symbol such as a rescaled length and wavenumber $(\tilde r,\tilde k)=(r/r_0,r_0 k)$, and a rescaled time and frequency coordinate $(\tilde t,\tilde\omega)=(\Omega t,\omega/\Omega)$. 

Furthermore, the perturbers of mass $M$ evolve on circular orbits in a background of FDM particles which they perturb through gravity. Following \cite{hui/etal:2017,Lan2020}, we can identify three characteristic velocities: the velocity dispersion $\sigma$ of the FDM particles, the orbital velocity $v_\text{circ}=\Omega r_0$ of the perturber, and a "quantum" velocity $v_Q\equiv GMm_a/\hbar$ where $m_a$ is the mass of the FDM particle. The latter can be interpreted as the perturber's escape velocity at a distance equal to the gravitational Bohr radius $(\hbar/m_a)^2/GM$. The velocities $\sigma$, $\Omega r_0$ and $v_Q$ can be used to define three distinct de Broglie wavelengths~:
\begin{align}
    \ls &\equiv \frac{\hbar}{m_a\sigma} \\
    &\simeq 1.918\times 10^{-2}\pc\, m_{18}^{-1} \left(\frac{\sigma}{100\kms}\right)^{-1} \nonumber \\
    \lv &\equiv \frac{\hbar}{m_a\Omega r_0} \\
    &\simeq 1.060\times 10^{-6}\pc\,m_{18}^{-1}\left(\frac{\Omega}{\yr^{-1}}\right)^{-1}\left(\frac{r_0}{\pc}\right)^{-1} \nonumber \\
    \lQ &\equiv \frac{\hbar}{m_a v_Q} \\
    &\simeq 854.1\pc\, m_{18}^{-2}\left(\frac{M}{M_\odot}\right)^{-1} \nonumber \;.
\end{align}
where $m_{18}\equiv m_a /10^{-18}\eV$.
$\ls$ and $\lv$ are the scales associated with the FDM velocity dispersion and the FDM-perturber relative velocity.
$\lQ$ can be thought of as the gravitational Bohr radius of the perturber \cite{Lan2020}. Using the orbit size (2$r_0$) as reference, these wavelengths imply three dimensionless ratios
\begin{align}
\rs&\equiv\frac{2r_0}{\ls} \\
&\simeq 104.3\, m_{18} \left(\frac{\sigma}{100\kms}\right)\left(\frac{r_0}{\pc}\right)\nonumber \\
\rv&\equiv \frac{2r_0}{\lv} \\
&\simeq 1.887\times 10^6\, m_{18}\left(\frac{\Omega}{\yr^{-1}}\right)\left(\frac{r_0}{\pc}\right)^2 \nonumber \\
\rQ &\equiv \frac{2r_0}{\lQ} \\
&\simeq 2.342\times 10^{-3}\,m_{18}^2 \left(\frac{M}{M_\odot}\right)\left(\frac{r_0}{\pc}\right) \nonumber \;.
\end{align}
They can be interpreted as a characteristic angular momentum (in unit of $\hbar$) associated with the FDM velocity dispersion, the relative velocity $\Omega r_0$ and the perturber's gravity. The larger their value the stronger the corresponding effect.
Their ratios lead to the Mach numbers $\ms=\rv/\rs$ and $\mQ=\rv/\rQ$ introduced in \cite{Lan2020}:
\begin{align}
    \ms &\equiv \frac{\Omega r_0}{\sigma} \\
    &\simeq 1.809\times 10^4 \left(\frac{\sigma}{100\kms}\right)^{-1}\left(\frac{\Omega}{\yr^{-1}}\right)\left(\frac{r_0}{\pc}\right) \nonumber \\
    \mQ &\equiv \frac{\Omega r_0}{v_Q} \\
    &\simeq 8.056\times 10^8\, m_{18}^{-1}\left(\frac{M}{M_\odot}\right)^{-1}\left(\frac{\Omega}{\yr^{-1}}\right)\left(\frac{r_0}{\pc}\right) \nonumber \;.
\end{align}
Our definition of $\mQ$ recovers that of \cite{Lan2020} in the linear motion case.

Finally, our treatment assumes linear response theory. Therefore, it is valid so long as fractional density perturbations are smaller than unity. For the FDM considered here, the scale $\lambda_\text{NL}$ below which nonlinearities are significant is obtained upon equating the "escape" velocity $v_Q$ with the circular velocity $\Omega r_0$:
\begin{align}
    \label{eq:Llin}
    \lambda_\text{NL} &\equiv \frac{2 v_Q}{\Omega} \\
    &\simeq 4.589\times 10^{-9}\pc\,m_{18}\left(\frac{M}{M_\odot}\right)\left(\frac{\Omega}{\yr^{-1}}\right)^{-1}\nonumber \;.
\end{align}
This scale naturally emerges from our calculations as the distance (from the perturber) at which the wake overdensity is of order unity (see Sec.~\S\ref{sec:denCon}).
$\lambda_\text{NL}\propto m_a$ because an increase in axion mass weakens the FDM "quantum pressure" that can oppose the gravitational pull.
Eq.~(\ref{eq:Llin}) should be contrasted to expression of $\lambda_\text{NL}\equiv GM/c_s^2$ in a gaseous medium with sound speed $c_s$, which is the Bondi radius \cite{Bondi:1952}.
Our analysis will be robust to nonlinear corrections provided that $r_0\gg \lambda_\text{NL}$, that is, $\mQ\gg 1$.

\begin{figure}
    \centering
    \hspace*{-1.8cm}
    \setlength\abovecaptionskip{-1.2\baselineskip}
    \input{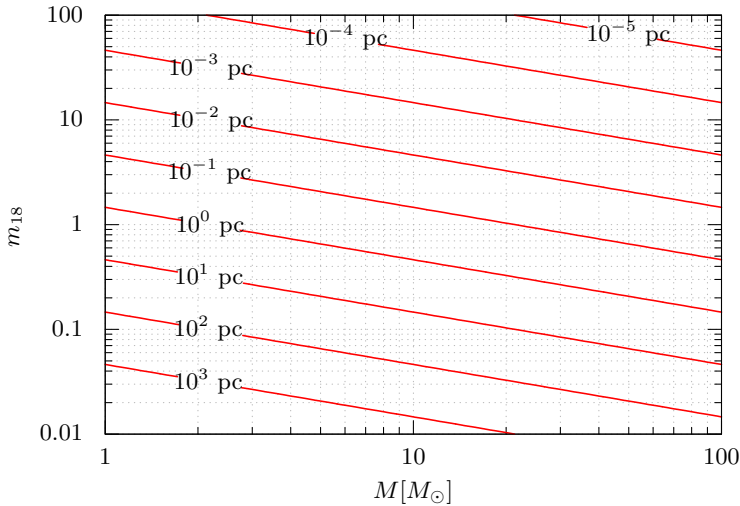}
    \caption{The orbital radius $r_0$ for which $\lambda_\text{NL}=0.1r_0$ (i.e. $\alpha_0$ defined in Eq.~\ref{eq:alpha0} is $\alpha_0=0.1$) is shown as a function of the total binary mass $M$ and the axion mass $m_{18}$. The nonlinear scale $\lambda_\text{NL}$ is calculated assuming circular Keplerian motion. The contour levels indicate the value of $r_0$ above which higher order contributions to the linear response approach considered here become larger than $\gtrsim 10$\%.
    }
    \label{fig:radiusNL}
\end{figure}

In order to gain insight into these parameters, let us assume $m_{18}=1$ and a FDM root-mean-square (rms) velocity dispersion $\sigma=100\kms$ appropriate to a Milky-Way size halo. $m_{18}=1$ will be our fiducial axion mass. This choice is motivated by recent astrophysical constraints (see Sec.~\ref{sec:introduction}).
Furthermore, we shall consider the following two configurations throughout this paper:
\begin{itemize}
    \item A single perturber of mass $1\msun$ on a circular orbit of radius $r_0=10^{-3}\pc$ around a super-massive black hole (SMBH) of mass $M_\bullet=10^6\msun$. The key parameters are
    \begin{align}
        \rs &\simeq 0.104 \\
        \rv &\simeq 3.95 \nonumber \\
        \lambda_\text{NL}&\simeq 2.19\times 10^{-9}\pc \nonumber \;,
    \end{align}
    whereas $\rQ\sim 10^{-6}$, $\mQ\sim 2\times 10^6$ and $\ms\sim 40$
    \item A compact, equal-mass binary of total mass $20\msun$ on a circular orbit of semi major axis $r_0=10^{-3}\pc$, in which case
    \begin{align}
        \label{eq:BinExample}
        \rs &\simeq 0.104 \\
        \rv &\simeq 1.77\times 10^{-2} \nonumber \\
        \lambda_\text{NL} &\simeq 4.90\times 10^{-6}\pc \nonumber \;.
    \end{align}  
    For this system, we have $\rQ\sim 10^{-5}$, $\mQ\sim 400$ and $\ms\sim 0.2$.
\end{itemize}
Note that, while $r_0\gg\lambda_\text{NL}$ in both cases, a globular cluster of mass $10^5\msun$ on a circular orbit of radius $1\Kpc$ in a dwarf galaxy halo of mass $10^8\msun$ implies a nonlinearity scale of $\lambda_\text{NL}\sim 20\Kpc$ much larger than the orbital radius. 

Fig.~\ref{fig:radiusNL} displays the orbital radius $r_0$ for which $\lambda_\text{NL}=0.1 r_0$ as a function of the total binary mass $M$ and the axion mass $m_{18}$. Contour levels indicate the $r_0$ above which higher order contributions to the linear response theory considered here roughly exceeds 10\%. This characteristic radius decreases with increasing axion mass because the "smoothing" from quantum pressure becomes weaker.

\section{Perturbing the GPP system}
\label{sec:GPP}

Fuzzy dark matter is in the form of a BEC and is thus described by a single wave function $\psi(\vr,t)$. Neglecting any possible self-interaction, the latter satisfies the (nonlinear) Schr\"odinger equation
\begin{equation}
  \label{eq:GP}
  i\partial_t\psi = - \frac{\hbar}{2m_a} \Delta_\vr\psi+\frac{m_a}{\hbar}\Phi\,\psi 
\end{equation}
where $\Phi$ is the gravitational potential and $\Delta_\vr \equiv \nabla_\vr^2 $ is the Laplacian. This notation makes clear that quantum mechanical effects appear only through the non-zero Compton length $\hbar/m_a$ of the particle \cite[see, e.g.,][for a discussion]{kaiser/widrow:1993,hui/etal:2017,mocz/lancaster/etal:2018}. 

Equation (\ref{eq:GP}) is supplemented by the Poisson equation
\begin{equation}
  \label{eq:P}
\Delta_\vr\Phi  = 4\pi G \rho\;,\qquad \rho \equiv |\psi|^2
\end{equation}
to form the Gross-Pitaievskii-Poisson (GPP) system.
The presence of a point-like perturber is included in the gravitational potential $\Phi=\Phi_0+\Phi_p$, which is the sum of the self-gravity $\Phi_0$ of the BEC and the potential $\Phi_p$ of the perturber. Namely, 
\begin{equation}
\label{eq:pertPot}
    \Delta_\vr\Phi_p = 4\pi G M\, h(t)\, \delta^D\!\big(\vr_p(t) - \vr\big)
\end{equation}
where $h(t)=1$ if the perturber is "turned on" and zero otherwise, and $\vr_p(t)=(r_0 \cos(\Omega t),\ r_0 \sin(\Omega t),\ 0)^T$ is the perturber's position. 

Two different routes can be taken to compute the linear response of the GPP system to a gravitational perturber: i) a wave scattering approach based on the Lippmann-Schwinger equation and ii) a sound propagation approach based on the Madelung form of GPP. Both are equivalent (so long as fully destructive interferences are absent).
The former can easily incorporate interference effects present in the atmosphere of FDM halos. However, it has a major drawback: the analytic calculation of DF is more challenging than in the hydrodynamic treatment. 

\subsection{Madelung hydrodynamic approach}
\label{sec:Madel}

To obtain the hydrodynamic form of the GPP system, we apply the Madelung transform \cite{madelung} 
\begin{equation}
    \label{eq:madt}
    \psi = \sqrt{\rho}\,\mathrm{e}^{i \theta}
\end{equation} 
where the phase $\theta$ is the velocity potential of a pure gradient flow
\begin{equation}
    \vv=\frac{\hbar}{m_a}\nabla_\vr\theta \;.
\end{equation}
Substituting Eq.~(\ref{eq:madt}) into the Schr\"odinger equation and extracting the real and imaginary part eventually leads to a continuity and momentum conservation equations reminiscent of ideal (non-viscous) hydrodynamics,
\begin{align}
    \label{eq:mad1}
    \partial_t\rho + \nabla_\vr(\rho \vv)&=0\\
    \partial_t \vv +(\vv\cdot \nabla_\vr)\vv &= -\nabla_\vr (Q +\Phi_0 +\Phi_p) 
    \nonumber \;.
\end{align}
Here, 
\begin{equation}
    \label{eq:Qpress}
    Q \equiv -\frac{\hbar^2}{2m_a^2}\frac{\Delta_\vr\sqrt{\rho}}{\sqrt{\rho}}\;.
\end{equation}
is the "quantum pressure" arising from the de-localization of the FDM particles.

\subsubsection{Linear "wave" equation}

On splitting the fluid variables into a (homogeneous) mean and a perturbation, $\rho\rightarrow \bar \rho +\delta \rho$ and $\vv \rightarrow \bar{\vv}+\delta\vv$, and linearizing Eqs~(\ref{eq:mad1}) 
in the velocity perturbation $\delta\vv$ and the fractional overdensity 
\begin{equation}
    \alpha(\vr,t)\equiv\frac{\rho(\vr,t)}{\bar \rho}-1 \;,
\end{equation}
we obtain
\begin{align}
    \label{eq:mandel1}
    \partial_t\alpha &=- (\bar{\vv}\cdot \nabla_\vr)\alpha -\nabla_\vr \delta\vv \\
    \partial_t \delta\vv &= -(\bar{\vv}\cdot \nabla_\vr)\delta\vv - \nabla_\vr \left(\Phi_0 +\Phi_p +\frac{\hbar^2}{4 m_a^2}\Delta_\vr \alpha\right)
    \nonumber \;.
\end{align}
Moving to the fluid rest frame ($\bar{\vv}=0$) and ignoring its self gravity $\Phi_0$, these equations reduce to a wave-like equation with a source term:
\begin{equation}
    \label{eq:WaveMadelung}
    \partial_t^2 \alpha +\frac{\hbar^2}{4m_a^2}\Delta_\vr^2 \alpha=-\Delta_\vr\Phi_p\;.
\end{equation}
Green's method can now be applied to compute the density contrast $\alpha$ and, thereby, the DF in linear response theory.

\begin{figure}[t]
    \centering
    \hspace*{-0.5cm}
    \includegraphics[width=.55\textwidth]{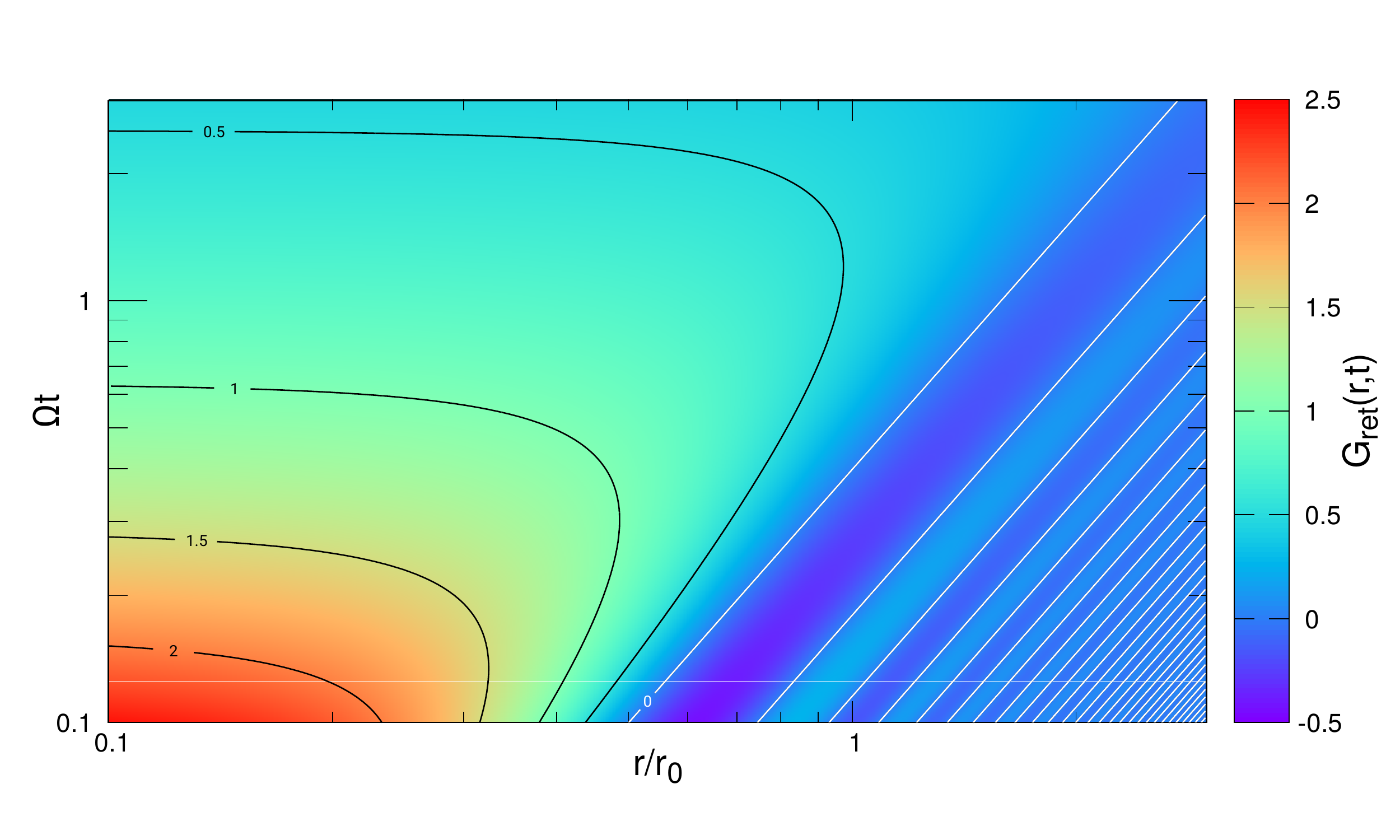}
    \caption{The retarded Green's function Eq.~(\ref{eq:greens}) assuming $\rv=4$. For convenience, $G_\text{ret}(\tvr,\tt)$ is normalized to $\frac{\rv}{4\pi \Omega r_0^3}$ and both $\tilde r$ and $\tilde t$ axes are logarithmic. The white contours indicate the locus for which $G_\text{ret}$ vanishes.}
    \label{fig:greens}
\end{figure}

\subsubsection{Green's function}

Transforming to the dimensionless variables $\tvk$, $\tomega$, the retarded Green's function for the wave equation (\ref{eq:WaveMadelung}) can be expressed as the Fourier transform 
\begin{equation}
    G_\text{ret}(\vr,\tau)=\frac{1}{\Omega r_0^3}\lim_{\epsilon\rightarrow0^+}\inttk \inttw \frac{e^{i (\tvk\cdot\tvr -\tomega\ttau)}}{\tk^4/\rv^2 -(\tomega+i \epsilon)^2}
    \label{eq:greens}
\end{equation}
This makes clear that $G_\text{ret}$ has dimensions of $[T L^{-3}]$ as it should.
Our shorthand notation is $\intw =\frac{1}{2\pi}\int_{-\infty}^{\infty} \mathrm{d}\omega$ and $\intk=\frac{1}{(2\pi)^3}\int_0^{2\pi}\mathrm{d}\varphi_k \int_{-1}^1 \mathrm{d} \cos(\vartheta_k)\int_0^\infty \mathrm{d}k\ k^2$ in the spherical coordinates used here (so that $\vk=(k,\varphi_k,\vartheta_k)$).
The condition $\epsilon>0$ (i.e. the poles lie in the lower half of the imaginary $\omega$-plane) enforces causality.

This integral can be solved analytically upon applying Cauchy's integral formula to the $\omega$ integration before performing the $\vk$ integral. We find
\begin{equation}
     \label{eq:greens}
    G_\text{ret}(\vr, t)=\frac{H(t)\,\rv}{4 \pi\tilde r\,\Omega r_0^3}\,\Im\!\left[\mathrm{erf}\left(\frac{1+i}{2\sqrt{2}}\sqrt{\rv} \frac{\tilde r}{\sqrt{\tilde t}}\right)\right]\;,
\end{equation}
where $\tilde r=|\tvr|$, $H(t)$ is the Heaviside function, $\Im(z)$ is the imaginary part of $z$ and $\mathrm{erf}(z)$ is the Error function. 

The Green's function is displayed in Fig.~\ref{fig:greens} as a function of $\tilde r$ and $\tilde t$ assuming $\rv=4$. 
In addition to the overall $1/\tilde r$ decrease of its amplitude, $G_\text{ret}$ oscillates around zero for large values of $\tilde r$. The onset of oscillations follows the relation $\tilde r/\sqrt{\tilde t}\sim$ const encoded in the functional dependence Eq.~(\ref{eq:greens}).
This is very different from the gaseous case, for which the Green's function $G_{\text{ret}}(\vr,t)=\frac{1}{r}\delta^D(t-r/c_s)$ is non-vanishing on the line $r=c_s t$ solely and cannot be negative. This forbids $\alpha<0$ and also leads to sharp discontinuities in the recovered density contrast unlike what is found in the FDM case (see for example Fig.~\ref{fig:alphaQMR4}). In the latter case, the oscillations and the absence of sharp features and high density caustics can be seen as a manifestation of the de-localized nature of FDM.

\subsubsection{Density contrast}
\label{sec:denCon}

Setting $\ttau=\tilde t-\tilde t'$ and $\tvu(\tilde t')=\tvr -\tvr_p(\tilde t')$ to be the dimensionless time difference and separation vector, respectively, the retarded Green's function (\ref{eq:greens}) yields the density contrast 
\begin{equation}
    \label{eq:alphaM}
    \alpha(\vr,t)=\alpha_0\int_0^{\infty}\mathrm d \tilde \tau\, \frac{h(\tilde t-\ttau)}{|\tvu(\tilde t-\ttau)|}\Im\!\left[\mathrm{erf}\left(\frac{1+i}{2\sqrt{2}}\sqrt{\rv} \frac{\tilde u}{\sqrt{\ttau}}\right)\right]
\end{equation}
where, again, $h(\tilde t)=1$ only when the perturber is turned on.
The normalization amplitude $\alpha_0$ is given by
\begin{equation}
    \label{eq:alpha0}
    \alpha_0\equiv \frac{2GM}{\lv(\Omega r_0)^2} = \frac{2}{\mQ} = \frac{\lambda_\text{NL}}{r_0}\;.
\end{equation}
It is identical to the normalization obtained in the case of a classical gas with negligible sound speed, for which the Bondi radius also is $GM/(\Omega r_0)^2$, except for $1/r_0$ being replaced by $1/\lv$. Assuming $\Omega$ independent of $r_0$, we have $\alpha_0\propto 1/r_0$ The value of $r_0$ for which $\alpha_0\equiv 1$ matches the nonlinearity scale Eq.~(\ref{eq:Llin}).

Our results readily extend to a binary with component masses $M_1=q_1 M$ and $M_2=q_2 M$ where $M=M_1+M_2$ is the total binary mass. The overdensity produced by the $j$-component ($j=1,2$) can be expressed as
\begin{multline}
    \label{eq:alphaMBinary1}
    \alpha_{j}(\tvr,\tilde t)=q_j\alpha_0\int_{0}^{\infty}\mathrm d \ttau\  \frac{h(\tilde t-\ttau)}{|\tvu_j(\tilde t-\ttau)|}\\\times\Im\left[\mathrm{erf}\!\bigg(\frac{1+i}{2\sqrt{2}}\sqrt{\rv} \frac{\tilde u_j}{\sqrt{\ttau}}\bigg)\right] \;,
\end{multline}
where $\tvu_{1}(\tilde t')=\tvr-q_{2}\tvr_p(\tilde t')$ and $\tvu_{2}(\tilde t')=\tvr-q_{1}\tvr_p(\tilde t'+\pi)$. The total overdensity is the sum of these two contributions:
\begin{equation}
    \label{eq:alphaMBinary2}
    \alpha(\tvr,\tilde t)=\alpha_1(\tvr,\tilde t)+\alpha_2(\tvr,\tilde t)\;.
\end{equation}
Results for both the single perturber and binary case are presented in Sec.~\S\ref{sec:wakes}.

\subsection{Lippmann-Schwinger approach}
\label{sec:LippSchw}

\begin{figure*}
    \centering
    \includegraphics[width=\textwidth, trim=0cm 4.5cm 0cm 4.5cm]{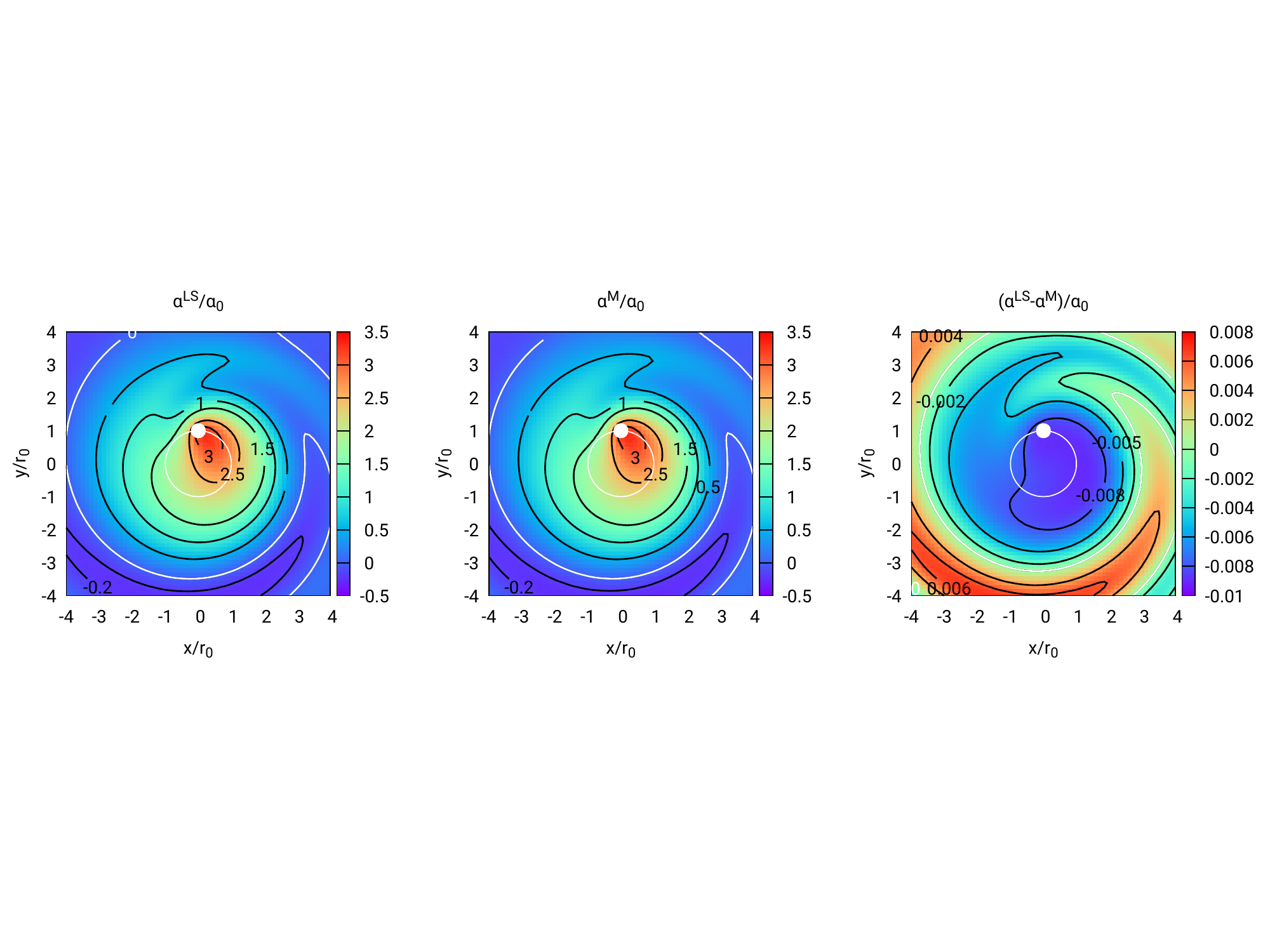}
    \caption{Density wake in the single perturber case computed after 1.25 rotations. The Born approximation to the Lippmann-Schwinger equation (left panel) and the linearized hydrodynamic approach (middle panel) are used. The corresponding overdensities $\als$ and $\am$ are normalized by $\alpha_0$ (see text for details).
    The right panel shows the difference $\als-\am$, which is at the few percents level. The parameter values $\rv=4$ and $\rs=0.1$ adopted here match our single perturber configuration.}
    \label{fig:alphaQMR4}
\end{figure*}

The Schr\"odinger equation Eq.~(\ref{eq:GP}) can also be recast into the form 
\begin{equation}
    \label{eq:scattering}
    \big(\hat E-\hat H\big)\psi=m_a \Phi \psi
\end{equation}
where $\hat E=i\hbar \partial_t$ and $\hat H=-\frac{\hbar^2}{2 m_a} \Delta_{\vec r}^2$ are operators, and $m_a\Phi$ can be treated as a (long-range) scattering potential. 

\subsubsection{Born approximation}

The Lippmann-Schwinger approach reformulates the solution $\psi$ of this scattering problem as an integral equation. In plain words, one writes
\begin{equation}
    \psi= \psi_0+\frac{1}{\hat E-\hat H}m_a \Phi \psi
\end{equation}
where $\psi_0$ is the homogeneous solution (free particle) to the Schr\"odinger equation. The Born series offers a recursive solution to this equation whereby, instead of searching for the full solution directly, (finite) successive approximations are produced iteratively starting from $\psi_0$:
\begin{align*}
    \psi_1&=\psi_0+\frac{1}{\hat E-\hat H}m_a \Phi \psi_0\\
    \psi_2&=\psi_1+\frac{1}{\hat E-\hat H}m_a \Phi \psi_1\\
    &\qquad\qquad\vdots
\end{align*}
Since we work in linear response theory throughout, we limit ourselves to the Born approximation (first order solution)
\begin{equation}
    \psi_1=\psi_0+\frac{1}{\hat E-\hat H}m_a \Phi \psi_0\equiv \psi_0+\delta\psi\;.
\end{equation}
We write the homogeneous solution as a superposition of plane waves \cite[e.g.][]{Bar2019}
\begin{equation}
    \label{eq:psi0}
    \psi_0(\vr,t)=\int_{\vk_0}\! \varphi(\vk_0)\, e^{i \vk_0\cdot\vr -i \omega_0 t}
\end{equation}
which fulfill the dispersion relation
\begin{equation}
    \label{eq:disp}
    \omega_0=\frac{\hbar k_0^2}{2 m_a} \;,
\end{equation}
while $\varphi(\vk_0)$ is a distribution to be determined later. In the first part of the following calculation, we will restrict ourselves to a monochromatic wave of arbitrary wavemode $\vk_0$ and normalized such that $|\psi_0|^2=\rho$ for simplicity. The distribution $\varphi(\vk_0)$ will be reintroduced below.

Our next focus is the (retarded) Green's function $(\hat E -\hat H)^{-1}$ which can be expressed as 
\begin{equation}
    \label{eq:GreensQM}
     G_\text{ret}(\vr,t)=\lim_{\epsilon\rightarrow 0^+}\intk \intw \frac{e^{-i \omega t+i \vk\cdot\vr}}{\hbar (\omega+i\epsilon)-\frac{\hbar^2}{2m_a} k^2}
\end{equation}
upon a Fourier transformation.
Finally, the last ingredient is the external potential $\Phi$ of the perturber, 
\begin{equation}
    \label{eq:Phi}
    \Phi(\vr,t)=-h(t)\frac{G M}{|\vr -\vr_p(t)|}\;.
\end{equation}
The self-gravity of the FDM component is, as before, neglected \cite[see, e.g.,][for treatments with the FDM self-gravity]{annulli/etal:2020,wang/easther:2022}

\subsubsection{Density contrast}

Substituting $G_\text{ret}$ and $\Phi$ into the Born approximation yields
\begin{equation}
    \delta \psi(\vr,t)=m_a\int\!\!\mathrm{d}r'^3\int \mathrm{d}t'\, G_\text{ret}(\vr-\vr',t-t')\, \Phi(\vr',t')\,\psi_0(\vr',t')
\end{equation}
This integral can be carried out with aid of the convolution theorem (further details are presented in Appendix~\ref{sec:LippSchw}).
In short, the Fourier transforms of $\Phi$ and $\psi_0$ can be combined with that of the Green's function (Eq.~\ref{eq:GreensQM}) to give
\begin{align}
    \delta \psi (\vr,t )&=-4 \pi m_a M G\int_{-\infty}^\infty\!\!\mathrm d \tau\, h(t-\tau) \psi_0(\vr_p(t-\tau),t-\tau) \nonumber \\ &\quad\times
    \lim_{\epsilon\rightarrow 0^+}\intk\intw\frac{e^{-i \omega\tau +i\vk\cdot\vu (t-\tau) }}{\hbar (\omega+i\epsilon) -\frac{\hbar^2 k^2}{2 m_a}}\frac{1}{|\vk-\vk_0|^2}
\end{align}
Performing first the integral over $\omega$ with help of Cauchy's integral formula leads to
\begin{multline}
    \label{eq:delPSi}
     \delta \psi(\tvr,\tilde t)=i \frac{\alpha_0}{2}\psi_0(r_0\tilde r,\tilde t/\Omega)\int_{0}^\infty\! \mathrm d \ttau\ \tilde h(\tilde t-\ttau)\\
     \times\frac{\mathrm{erf}\big(\frac{1-i}{2}\sqrt{\frac{\rv}{2\ttau}} |\tvu(\tilde t-\ttau)-\frac{2}{\rv}\ttau \tvk_0|\big)}{|\tvu(\tilde t-\ttau)-\frac{2}{\rv}\ttau \tvk_0|}
\end{multline}
which is expressed in terms of the dimensionless variables. 

Our calculation is thus far limited to a monochromatic wave, but it can be readily extended to any superposition of plane waves with arbitrary amplitudes $\varphi(\tvk_0)$ along the lines of \cite{hui/etal:2017,Bar2019,Lan2020}.
Incoherent superposition of FDM granules or wave packets occurs, for instance, inside virialized halos \cite{schive/etal:2014a,schive/etal:2014b,schwabe/etal:2016}.
Interpreting the group velocity of (almost monochromatic) wave packets
$\vv=\hbar\tvk_0/m r_0$ as the velocity of FDM quasi-particles, we draw the amplitudes $\varphi(\tvk_0)$ from a Maxwell-Boltzmann distribution 
\begin{equation}
   f(\tvk_0)= \left(\frac{2}{\pi}\right)^{3/2}\frac{\rho}{\rs^3}\,
   e^{-\frac{2\tvk_0^2}{\rs^2}}
   \label{eq:MxBzDis}
\end{equation}
normalized such that 
\begin{equation}
    \big\langle|\psi_0|^2\big\rangle=\int_{\tvk_0}\! f(\tvk_0)=\rho \;.
\end{equation}
Assuming that the amplitudes $\varphi(\tvk_0)$ are statistically independent, we require
\begin{equation}
    \big\langle\varphi(\tvk_0)\varphi^*(\tvk'_0)\big\rangle=f(\tvk_0)\, \delta^D (\tvk_0-\tvk'_0)\;.
\end{equation}
The ensemble average is performed over realizations of FDM backgrounds with prescribed velocity dispersion $\sigma$ and spatial average density $\rho$.

Introducing $\delta\psi'=\delta\psi/\psi_0$, the ensemble average density contrast is now calculated as
\begin{align}
    \langle\alpha\rangle
    &=\frac{1}{\rho}\big\langle |\psi|^2\big\rangle-1 \nonumber \\
    &=\frac{1}{\rho} \big\langle|\psi_0|^2\big\rangle\,|1+\delta\psi'|^2-1 \nonumber\\
    \label{eq:alphaQMRaw}
    &=\frac{1}{\rho}\int_{\tvk_0}\! f(\tvk_0)\, 
    \big(2\Re{(\delta\psi')}+|\delta \psi'|^2\big) \;.
\end{align}
Here again, the result for the single perturber case easily extends to a compact binary upon defining
\begin{multline}
    \label{eq:alphaLSBinary1}
    \delta \psi_j'(\tvr,\tilde t)=i\ q_j\frac{\alpha_0}{2}\int_{0}^\infty \mathrm d \ttau\ \tilde h(\tilde t-\ttau)\\  \times
    \frac{\mathrm{erf}\big(\frac{1-i}{2}\sqrt{\frac{\rv}{2\ttau}} |\tvu_j(\tilde t-\ttau)-\frac{2}{\rv}\ttau \tvk_0|\big)}{|\tvu_j(\tilde t-\ttau)-\frac{2}{\rv}\ttau \tvk_0|} 
\end{multline}
This allows us to express the average density wake produced by a compact binary as
\begin{multline}
    \label{eq:alphaLSBinary2}
    \langle\alpha\rangle=\frac{1}{\rho}\int_{\tvk_0}\! f(\tvk_0)\,\big(2\Re(\delta \psi'_{1})+2\Re(\delta \psi'_{2})+|\delta \psi'_{1}|^2+|\delta \psi'_{2}|^2\\+\delta\psi'_{1}\delta\psi^{\prime*}_2+\delta\psi_{1}^{\prime*}\delta\psi_2'\big)\;.
\end{multline}
A Maxwell-Boltzmann distribution for $f(\tvk_0)$ will be assumed like in the single perturber case.

\begin{figure*}
    \centering
    \includegraphics[width=\textwidth, trim=0cm 4.5cm 0cm 4.5cm]{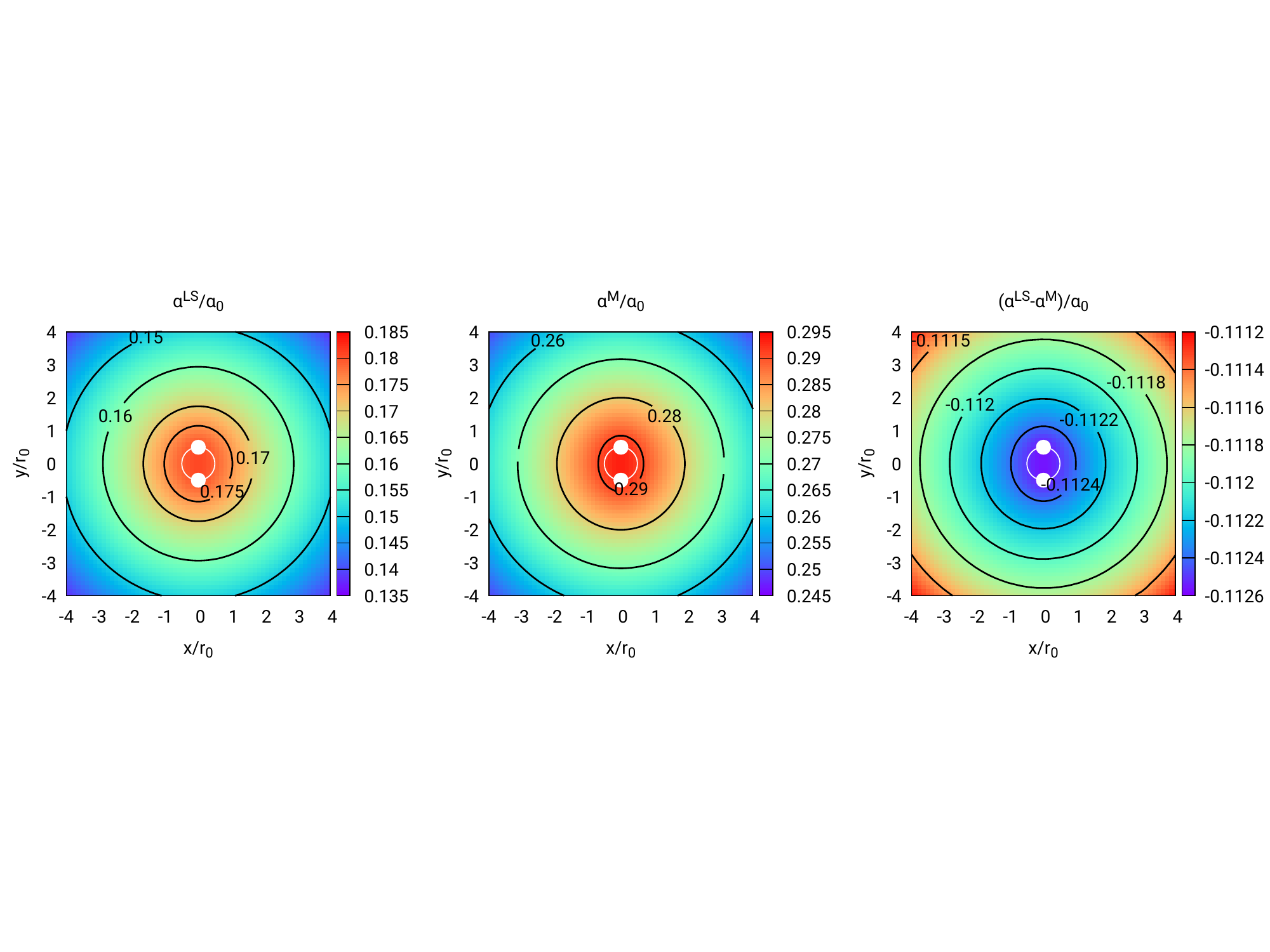}
    \caption{Similar to Fig.~\ref{fig:alphaQMR4} but for our fiducial compact binary with mass ratios $q_1=q_2=0.5$, $\rv=0.0177$ and $\rs=0.1$.
    The predicted wave- and fluid-like overdensities $\als$ and $\am$ are normalized to $\alpha_0$ (see text for details).
    The right panel shows the difference $\als-\am$.
    The current position of the binary components are indicated by white filled symbols, while the binary orbit is shown as a thin white circle. 
    }  
    \label{fig:alphaQMR01}
\end{figure*}

\subsection{Density wakes}
\label{sec:wakes}

\subsubsection{Numerical implementation}
\label{sec:numerics}

The density wakes $\alpha(\vr,t)$ presented in this Section are produced by numerically solving the initial value problem in the Madelung and Lippmann-Schwinger approach on a regular, 2-dimensional mesh of $64\times 64$ grid points covering the domain $[-4r_0,+4r_0]\times [-4r_0,+4r_0]$. 

In the hydrodynamic approach, we use an initially homogeneous density distribution which we evolve according to Eq.~(\ref{eq:alphaM}) (single perturber) and to Eqs.~(\ref{eq:alphaMBinary1}) -- (\ref{eq:alphaMBinary2}) (compact binary). In the wave approach, we solve Eq.~(\ref{eq:alphaQMRaw}) (single perturber) and Eq.~(\ref{eq:alphaLSBinary2}) (compact binary) after the substitution of Eq.~(\ref{eq:delPSi}).

These numerical results thus correspond to the "finite time perturbation" case and, as we will see below, never achieve steady-state. We discuss the convergence to steady-state further in Sec.~\S\ref{sec:finti} and \S\ref{sec:NumCom} when we explore DF. For the latter, sampling along the $z$-axis is also needed. We consider a $64^3$ cubical grid of size $8r_0$ although, in practice, we take advantage of the planar symmetry and add 32 evenly distributed points in the interval $[-4 r_0, 0]$ along the $z$-direction. The comparison between DF computed from these numerical results and from our analytical expressions can be found in Sec.~\ref{sec:NumCom}.

All our simulations have absorbing boundary conditions to prevent the artificial reflection of the density wake. 

\subsubsection{Results}

Fig~\ref{fig:alphaQMR4} displays the density wake created by our fiducial single perturber after it is turned on at $\tilde t=0$. Results are shown both in the Lippmann-Schwinger approach ($\als$, left panel) and in the Madelung approach ($\am$, middle panel). 
The density contrast $\alpha(\vr,t)$ is plotted in the orbital plane after the completion of $1.25$ rotations. The white symbol indicates the perturber's position on its circular orbit. The wake produced by the perturber's gravitational disturbance is a deformed ellipsoid in the vicinity of the circular orbit. The inner overdense wake is surrounded by an underdense, outwardly spiraling whose tip slightly lags behind the perturber. The main differences with the wake pattern in the gaseous case \cite[see for instance][]{kim/etal:2007,kim/etal:2008,DesNuBu21} are the existence of underdense regions together with the absence of sharp discontinuities and caustics (the latter arise in the gaseous case when the motion is supersonic). 
In the FDM case, small scale density features are smoothed out by the "quantum pressure" Eq.~(\ref{eq:Qpress}), reflecting the de-localized nature of the FDM particles.

Likewise, the density wake for the compact binary case shown in  Fig.~\ref{fig:alphaQMR01} is smoothly distributed around the binary center of mass, with a slight elliptic asymmetry aligned with the position of the bodies. The outer, underdense spiraling region visible in the single perturber case is not present for the equal mass binary considered here because the two bodies produce a spiral of equal size in a symmetric fashion, which adds up to a circular distribution. 
Increasing $\rv$ away from the fiducial value $\rv=0.0177$ lessens the smoothing due to $\lv$ and produces (weak) deformations of the wake's circular shape, albeit nothing comparable to the spirals seen in the single perturber case. As expected, unequal mass binaries produce more asymmetric distributions. 

The prefactor $\alpha_0=\lambda_\text{NL}/r_0$ controls the convergence of the perturbative solution both in the Madelung and in the Lippmann-Schwinger approach. As a result, the contribution proportional to $|\delta \psi'|^2$ in Eq.~(\ref{eq:alphaQMRaw}) formally is second order. For our choice of fiducial parameters that ensures $\alpha_0\ll 1$, it is negligible relative to the first order term. Therefore, we have discarded these second order terms in the calculation of $\als$. 

The right panel of figs~\ref{fig:alphaQMR4} and \ref{fig:alphaQMR01} shows the difference between the overdensity $\als$ and $\am$ computed in the wave and hydrodynamic approach, respectively. Our Madelung prediction $\am$ is equal to $2\Re{(\delta\psi')}$ evaluated at $\vk_0=0$. Therefore, it misses the contribution $\frac{2}{\rv}\tvk_0\ttau=\frac{\hbar}{m_a}\vk_0\tau$ to the separation vector in Eq.~(\ref{eq:delPSi}). This additional term is the distance traveled by a FDM quasi-particle with group velocity $\frac{\hbar}{m_a}\vk_0$ in the time interval $\tau$.
The Maxwell-Boltzmann distribution implies that the relevant wavenumbers satisfy $\tilde k_0\lesssim \rs/2$. This defines a characteristic scale $(\rs/\rv) \ttau=\ttau/\ms$ over which the FDM quasi-particles are redistributed and the wake density contrast is smoothed.

The impact of the FDM velocity dispersion is small in the single perturber case because $\ms\gg 1$ for our fiducial choice of parameters. The density wake $\als$ computed with a Maxwell-Boltzmann distribution has an amplitude lower by about 0.2\% in overdense regions relative to $\am$, and larger by at most 3\% in underdense regions. 
By contrast, the difference between $\als$ and $\am$ is markedly stronger in the binary case since we now have $\ms\ll 1$. In the center where the overdensity is highest, the FDM velocity dispersion lowers the wake amplitude by roughly 38\%. This suppression reaches up to 42\% in the outer region of lower (but still positive) overdensity.

\section{Dynamical friction}
\label{sec:DF}

Our calculation of DF follows the approach outlined in \cite{DesNuBu21} which we shall briefly summarize to begin with.

\subsection{Multipole expansion}

The dynamical friction can be expressed as 
\begin{equation}
    \label{eq:dynFricRaw}
    \mathbf F_{\mathrm {DF}}(t)=G M \rho\int\mathrm{d}^3u\, \frac{\vu}{u^3}\,\alpha(\vu,t)\;,
\end{equation}
where $\vu$ is the separation vector between the current position $\vr_p=\vr_p(t)$ of the perturber and the wavefront produced by it from an earlier position $\vr_p'=\vr_p(t')$ at the retarded time $t'<t$. For the FDM medium considered here, the overdensity $\alpha(\vu,t)$ is computed either in the Madelung or in the Lippmann-Schwinger approach as discussed above. 

The hydrodynamic formulation has the advantage that, in linear response theory, $\alpha$ is given by the convolution Eq.~(\ref{eq:greens}) of the source with the Green's function like in the gaseous case explored in \cite{DesNuBu21}. On expanding part of the complex exponential $e^{i\tvk\cdot\tvr}$ (here, $\tvr = \frac{1}{r_0}(\vr_p-\vr_p'+\vu$)) in plane waves, decomposing the force into a helicity basis and performing the integral over the orientation of $\tvk$, the DF can be re-expressed as 
\begin{equation}
    \mathbf F_{\mathrm{DF}}(t)=-4 \pi \left(\frac{G M}{\Omega r_0}\right)^2\rho\, \big(\Re(I)\,\hvr(t)+\Im(I)\,\hpv(t)\big)
\end{equation}
where $\hvr(t)$ and $\hpv(t)$ are unit vectors in the radial and tangential direction, respectively. $\Re(I)$ and $\Im(I)$ are the real and imaginary part of a dimensionless function $I=I(\rv)$ which, on exploiting symmetry relations of Wigner 3$j$ symbols, can be recast into the form
\begin{align}
    \label{eq:Im}
    I &=\sum_{l=1}^{\infty}\sum_{m=-l}^{l-2} (-1)^m \frac{(l-m)!}{(l-m-2)!}\\
    &\qquad\times\frac{\sllm(m,\rv,t)-\sllm^{\ *}(m+1,\rv,t)}{\Gamma(\frac{1-l-m}{2})\Gamma(1+\frac{l-m}{2})\Gamma(\frac{3-l+m}{2})\Gamma(1+\frac{l+m}{2})} \nonumber\;.
\end{align}
Here, $\Gamma(z)$ is the usual Gamma function while the "scattering amplitude" 
\begin{multline}
    \label{eq:scat}
    \sllm(m,\rv,t)=\lim_{\epsilon\rightarrow0^+}\inttw \int_{-\infty}^{\infty}\mathrm{d}\ttau\ h(\tilde t-\ttau)\, e^{i(m -\tilde\omega)\ttau}\\ \times\int_0^\infty \dtk\  \frac{\tk\,\bessJ{l}{\tk}\,\bessJ{l-1}{\tk}}{\tk^4/\rv^2-(\tilde \omega+i\epsilon)^2}
\end{multline}
involve products of radial standing waves (spherical Bessel functions) $\bessJ{l}{z}$.
The sign of $\Re(I)$ and $\Im(I)$ determines whether the radial DF force is a weight ($\Re(I)>0$) or a lift ($\Re(I)<0$), and the tangential DF force a drag ($\Im(I)>0$) or a thrust ($\Im(I)<0$).

\begin{figure}[t]
    \centering
    \setlength\abovecaptionskip{-0.7\baselineskip}
    \hspace*{-0.7cm}
    \input{single_pert_comp_kmin_0.3}
    \caption{The real and imaginary part of the function $I(\rv)$ (solid curves).
    in FDM backgrounds calculated from Eq.~(\ref{eq:Im}) with (\ref{eq:slls}) to (\ref{eq:styM0}) with $\tkmin=0.3$. For comparison, we also show the corresponding gaseous $I(\mg)$ (dashed curves) computed in \cite{DesNuBu21}.
    Although the quantity $\rv$ plays a role similar to the gas Mach number $\mg$, we show $I(\rv)$ as a function of $\sqrt{\rv}$ for convenience. Furthermore, we have rescaled the real and imaginary parts of $I$ such that they all asymptotes to a constant in the limit of large $\rv$ or $\mg$. For the imaginary part of $I(\rv)$, this rescaling leads to an artificial divergence at $\rv=0$. The multipoles have been summed up to $l_\mathrm{max}=100$ in both the gas and FDM case.}
    \label{fig:fdmGasComp}
\end{figure}

A comparison with \cite{hui/etal:2017,Lan2020} shows that their friction coefficient $\mathcal{C}$ is equivalent to our $\Im(I)$. In the circular case, $I$ is a "complex friction" which encodes also the lift/weight in the radial direction.

\subsection{Single perturber} 

\subsubsection{Steady state}
\label{sec:sty}

For a perturber in steady state (labeled with "Sty"), $h(t)=1$ and the integral of $\tau$ in Eq.~(\ref{eq:scat}) returns $2\pi \delta^D(m\Omega-\omega)$ independently of $t$. This can be used to simplify the scattering amplitude further to
\begin{equation}
    \label{eq:sllStart}
    \slls(m,\rv)=\lim_{\epsilon\rightarrow0^+}\int_0^\infty \dtk\  \tk \frac{\bessJ{l}{\tk}\bessJ{l-1}{\tk}}{\tk^4/\rv^2-(m+i \epsilon)^2}
\end{equation}
This integral can be solved upon expressing the spherical Bessel functions in terms of the Hankel functions $\hone{l}{z}$ and $\htwo{l}{z}$ and applying Cauchy's integral formula. 
Details of the calculation can be found in Appendix~\ref{sec:detStdy}.
The result can be compressed further with the aid of several identities of spherical Bessel functions, and eventually recast into
\begin{multline}
    \slls(m,\rv)=\frac{i \pi \rv}{4m}\Big[\bessJ{l}{\sqrt{ m \rv}}\,\hone{l-1}{\sqrt{ m \rv}}\\+\frac{i}{\sqrt{ m \rv}}\textit{I}_{l+1/2}\big(\sqrt{ m \rv}\big)\textit{K}_{l-1/2}\big(\sqrt{ m\rv}\big)\Big]\;.
    \label{eq:slls}
\end{multline}
This expression is strictly valid for azimuthal numbers $m\neq0$ only. When $m=0$, the integrand of Eq.~(\ref{eq:sllStart}) exhibits a single pole at $k=0$ and a very different behavior depending on the value of $l$. When $l\neq1$, the scattering amplitude $\slls(0,\rv)$ is finite and equal to
\begin{equation}
    \slls(0,\rv)=\frac{3 \pi \rv^2}{18-80l^2+32l^4}\;.
    \label{eq:slls0}
\end{equation}
When $l=1$, the purely real amplitude $S^{\mathrm{Sty}}_{1,0}(0,\rv)$ has an infrared divergence which we regularize by introducing a lower cut-off $\tkmin$:
\begin{align}
    \label{eq:styM0}    
    S^{\mathrm{Sty}}_{1,0}(0,\rv)
    &=\rv^2\int_{\tkmin}^\infty\frac{\dtk}{\tk^3} \bessJ{1}{\tk}\bessJ{0}{\tk} \\
    &=\frac{\rv^2}{40 \tkmin^5}\Big[ 4-4\pi\tkmin^5+\big(4\tkmin^4-2\tkmin^2-4\big) \nonumber \\
    &\quad\times\cos(2\tkmin)+\tkmin\big(2\tkmin^2-3\big)\sin(2\tkmin) 
    \nonumber \\
    &\quad +8\tkmin^5\mathrm{Si}(2\tkmin)\Big]
    \nonumber \;.
\end{align}
Here, $\mathrm{Si}(z)$ is the sine-integral.
The previous expressions are all that we need in order to compute the "complex friction" $I(\rv)$ in the steady-state regime.

\begin{figure}
    \centering
    \hspace*{-0.7cm}
    \setlength\abovecaptionskip{-0.7\baselineskip}
    \input{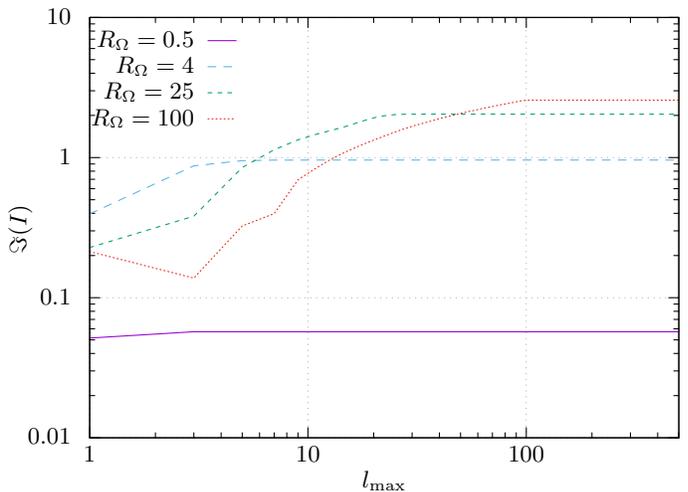}
    \caption{The imaginary part $\Im(I)$ of the "complex friction" $I(\rv)$ is shown for different values or $\rv$ as a function of the the upper limit $l_{\mathrm{max}}$ of the (truncated) multipole expansion Eq.~(\ref{eq:Im}). Unlike the gaseous case, $\Im(I)$ converges to any desired accuracy for finite values of $l_{\mathrm{max}}$.}
    \label{fig:conv}
\end{figure}

In Fig.~\ref{fig:fdmGasComp}, we compare the steady-state FDM $I(\rv)$ predicted by the Madelung approach with the corresponding gaseous solution $I(\mg)$ given in \cite{DesNuBu21}. The FDM parameter $\rv$ plays a role similar to the gas Mach number $\mg$, at least in the way it appears in the scattering amplitude 
(compare the $k$-integrand of Eq.~(\ref{eq:sllStart}) with that of Eq.~(17) in \cite{DesNuBu21}).
Nevertheless, we have plotted $I(\rv)$ as a function of $\sqrt{\rv}$ so that $\Im(I(\rv))$ approximately reaches its maximum where $\Im(I(\mg))$ does.
In the gaseous case, we normalize the real part $\Re(I(\mg))$ such as to emphasize that it asymptotes to $\mg^2$ in the limit of large Mach numbers. We use the same normalization for $\Im(I(\mg))$. In the FDM case, $\Im(I(\rv))$ is fully determined by $\slls(m,\rv)$ with $m\neq0$, which depends linearly on $\rv$.  However, $\Re(I(\rv))$ also depends on $\slls(0,\rv)$, which is quadratic in $\rv$. This leads to a linear dependence at small $\rv$ and a quadratic one at large $\rv$. Showing $\Re(I(\rv))/\rv^2$ as done in the figure filters out the large-$\rv$ asymptotic, and leads to the artificial divergence seen at small $\rv$. 

The quantity $\Im(I(\rv))$ reaches a maximum for $\rv\sim\mathcal{O}(1)$ around which the synchronization between the perturber and its wake is most efficient. In the gaseous case, this occurs at $\mg=1$.
However, $I(\rv)$ exhibits less features relative to $I(\mg)$ owing to the quantum pressure.
Another important difference can be spotted upon comparing our Fig.~\ref{fig:conv} to Fig.~3 of Ref.~\cite{DesNuBu21}. Both of them display the dependence of the imaginary part $\Im(I)$ (for a few different values of $\rv$ and $\mg$, respectively) on the upper limit $l_{\mathrm{max}}$ of the sum defining $I$ in Eq.~(\ref{eq:Im}). 
While the gas case exhibits a logarithmic divergence for supersonic Mach number, there is no such behavior for FDM. The $k^4$-scaling of the (Fourier space) Greens' function ensures the convergence of the multipole expansion regardless of the value of $\rv$. In Fig.~\ref{fig:conv}, the series convergence is achieved for $l_{\mathrm{max}}\gtrsim 100$.
We also checked the real part assuming $\tkmin=0.3$ and found even faster convergence. For $\rv=100$ for instance, $\Re(I)$ has already converged by $l_{\mathrm{max}}\approx 20$.

The different behavior of the gaseous $I(\mg)$ and FDM $I(\rv)$ is related to the features seen in Fig.~\ref{fig:greens} and Fig.~\ref{fig:alphaQMR4} and discussed at the end of Sec.~\S\ref{sec:Madel}. The quantum pressure prevents the FDM density wake to form jump discontinuities whereas, in a gaseous medium, there are sharp discontinuities at which the density increases towards the perturber. They lead to the short-distance (ultraviolet) divergence of $\Im(I(\mg))$ for point-like perturbers \cite[see][]{kim/etal:2007,DesNuBu21}.

\begin{figure}
    \centering
    \setlength\abovecaptionskip{-0.7\baselineskip}
    \hspace*{-0.9cm}
    \input{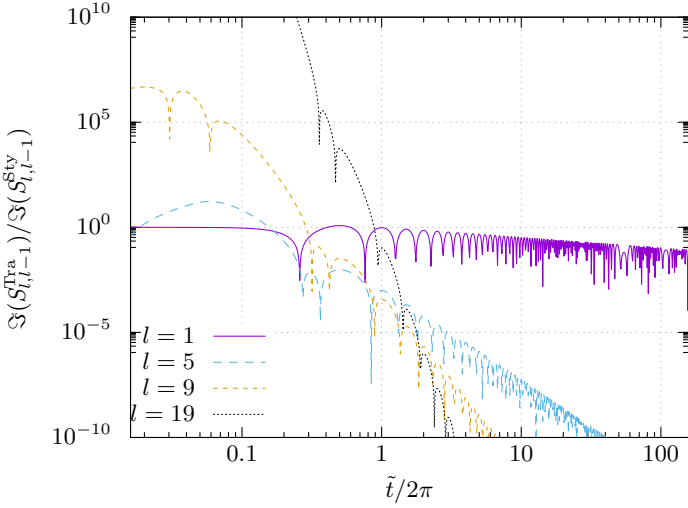}
    \caption{The imaginary part of the transient amplitude $\sllt$ normalized to its steady-state counterpart as a function of the (dimensionless) time $\tilde t$. Results are shown for a few choices of multipole $l$. The azimuthal number is fixed to $m=1$ and a parameter value $\rv=4$ is assumed. All the transient amplitudes shown here fall below the steady-state result after completing one rotation. Their envelope approximately decays as $\sllt\propto t^{-l+1/2}$.}
    \label{fig:trans}
\end{figure}

\subsubsection{Finite time perturbation}
\label{sec:finti}

The "finite time perturbation" (labeled with "Ftp") corresponds to a perturber "turned on" at time $t=0$ and can be straightforwardly explored with numerical simulations. Details of the derivation are given in Appendix~\ref{sec:detFin}. In short, $h(t-\tau)=0$ for $\tau>t$ such that the integral over $\ttau$ becomes
\begin{equation}
    \int_{-\infty}^{\tilde t}\!\mathrm{d}\ttau\ e^{i (m-\tilde \omega)\ttau}=\lim_{\eta\rightarrow0^+}\frac{e^{i (m-\tilde\omega)\tilde t}}{i\big(m-\tilde \omega-i \eta\big)}\;.
    \label{eq:transStart}
\end{equation}
The resulting scattering amplitude can be written as $\sllf(m,\rv,t)=\slls(m,\rv)+\sllt(m,\rv,t)$, where the transient amplitude (labeled with "Tra") is obtained upon taking the limit $\epsilon\to 0^+$ of
\begin{multline}
    \sllt(m,\rv,t)=-\frac{\rv}{2}e^{i m\tilde t}\int_0^\infty \frac{\dtk\ }{\tk} \bessJ{l}{\tk}\bessJ{l-1}{\tk}\\\times\bigg(\frac{e^{-i( \tk^2/\rv-i \epsilon)\tilde t}}{\tk^2/\rv-m -i\epsilon}+\frac{e^{i(\tk^2/\rv+i \epsilon)\tilde t}}{k^2/\rv+m +i\epsilon}\bigg)\;.
    \label{eq:transM0}
\end{multline}
Notice that we have $\sllt(m,\rv,0)=-\slls(m,\rv)$ at $t=0$, which ensures that the DF initially vanishes.

For $(l,m)=(1,0)$, Eq.~(\ref{eq:transM0}) is purely real. Like the corresponding steady-state $S_{1,0}^\text{Sty}$, the transient $S_{1,0}^\text{Tra}$ exhibits an infrared divergence which can be again remedied by introducing a lower cut-off $\tkmin$ in the $\tk$-integration. However, we have not been able to find an analytic expression for $\tilde t >0$.

For $(l,m)\neq (1,0)$, we can apply Cauchy's integral formula and rewrite Eq.~(\ref{eq:transM0}) as 
\begin{align}
    \sllt&=-\frac{\rv}{2}e^{i m\tilde t}\int_0^\infty\frac{\mathrm{d}\tilde\chi}{\tilde \chi}\big[ \bessJ{l}{(1+i)\tilde \chi}\bessJ{l-1}{(1+i)\tilde \chi} \nonumber \\ &\qquad -\bessJ{l}{(1-i)\tilde \chi }\bessJ{l-1}{(1-i)\tilde \chi}\big]\nonumber \\ & \qquad \times\frac{e^{-2\tilde t \tilde\chi^2/\rv}}{2i \tilde \chi^2/\rv+m +i\epsilon}
    \label{eq:sllt} \;,
\end{align}
which makes clear that the transient contributions $\sllt$ decay so long as $(l,m)\neq (1,0)$.

\begin{figure}
    \centering
    \setlength\abovecaptionskip{-0.7\baselineskip}
    \hspace*{-0.9cm}
    \input{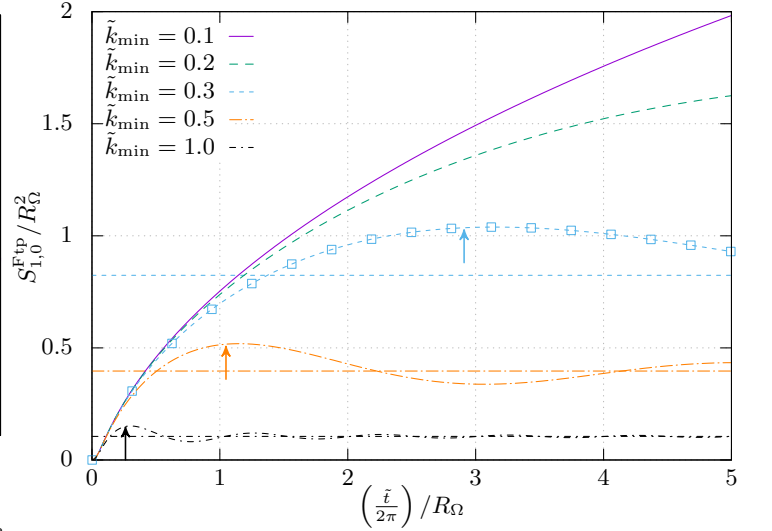}
    \caption{The finite time perturbation amplitude $\sllf$ in the special case $l=1$ and $m=0$. Results are shown as a function of the number of rotations $\tilde t/2\pi$ (in unit of $\rv$) for different lower cut-off wavenumbers $\tkmin$ as indicated on the figure. All the curves assume $\rv=4$, while the empty symbols represent the particular case $(\tkmin,\rv)=(0.3,1)$. The (also) diffusive nature of the Schr\"odinger equation leads to the square root behavior $S_{1,0}^\text{Ftp}\propto\sqrt{\tilde t}$ at early time. Arrows mark the estimated timescale at which the initial ($t=0$) perturbation has diffused throughout the medium (see text). The horizontal lines indicate the steady-state limit which the finite time perturbation result converges to.}
    \label{fig:transM0}
\end{figure}

Numerical evaluations of $\sllt$ are shown in Fig.~\ref{fig:trans} as a function of the dimensionless time $\tilde t$ for a fixed azimuthal number $m=1$ but different choices of multipole $l$. A value of $\rv=4$ is assumed. The transient amplitudes shown here are normalized to their steady-state counterpart $\slls$ to facilitate the comparison. Their envelope decays with time according to (the empirical law) $\sllt\propto t^{-l+1/2}$. Consequently, with the notable exception of $S_{1,0}^\text{Tra}(0,\rv,t)$, essentially all the transient contributions drop below the steady-state result after a rotation at most, regardless of the value of $l$. Consequently, $\sllf\approx \slls$ at better than a percent level after a few rotations solely.

Nonetheless, neither the radial nor the tangential component of the finite perturbation time DF ever reaches the steady-state regime.
In the radial direction, the convergence depends on $\Re(I)$ and, therefore, is slowed down by the infrared divergence present for $(l,m)=(1,0)$. This is emphasized in Fig.~\ref{fig:transM0}, in which we plot the purely real $S^\text{Ftp}_{1,0}(0,\rv,t)$ (in unit of $\rv^2$) as a function of the number of rotations $\tilde t/2\pi$ (in unit of $\rv$). All the curves assume the same $\rv=4$ but a different $\tkmin$ as labeled on the figure. The empty symbols show results for $\rv=1$ in the particular case $\tkmin=0.3$, while the horizontal lines indicate the steady-state solution.
For a cut-off wavenumber as small as $\tkmin=0.1$, $S^\text{Ftp}_{1,0}(0,\rv,t)$ does not (even) reach the magnitude of the steady-state solution after 20 rotations. For larger  $\tkmin\gtrsim 0.3$ however, convergence to steady-state occurs faster.

\begin{figure*}
    \setlength\abovecaptionskip{-0.7\baselineskip}
    \begin{minipage}{1.\textwidth}
        \includegraphics[width=1\textwidth,height=8.5cm]{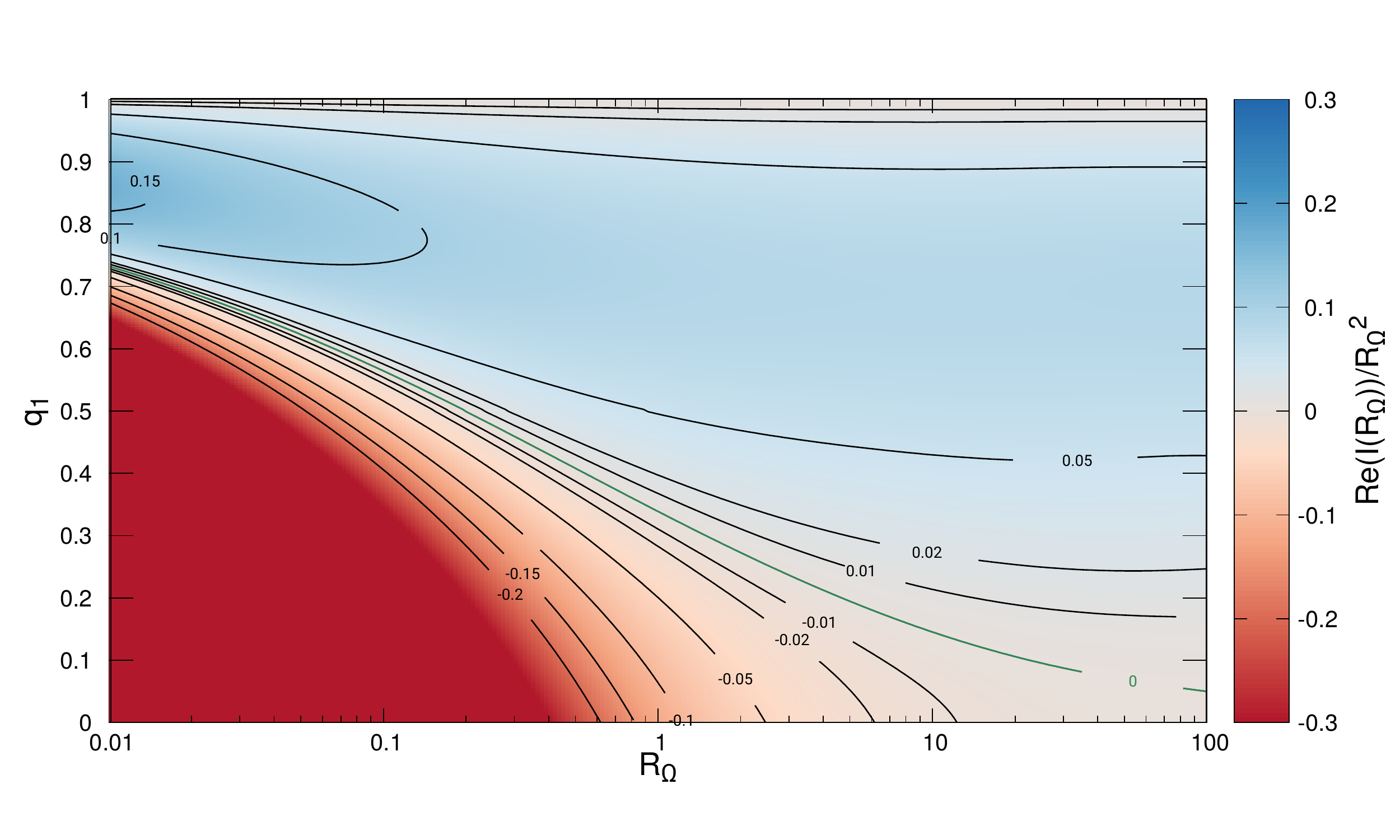}  
    \end{minipage}
    \hfill
    \begin{minipage}{1.0\textwidth}
        \includegraphics[width=1\textwidth,height=8.5cm]{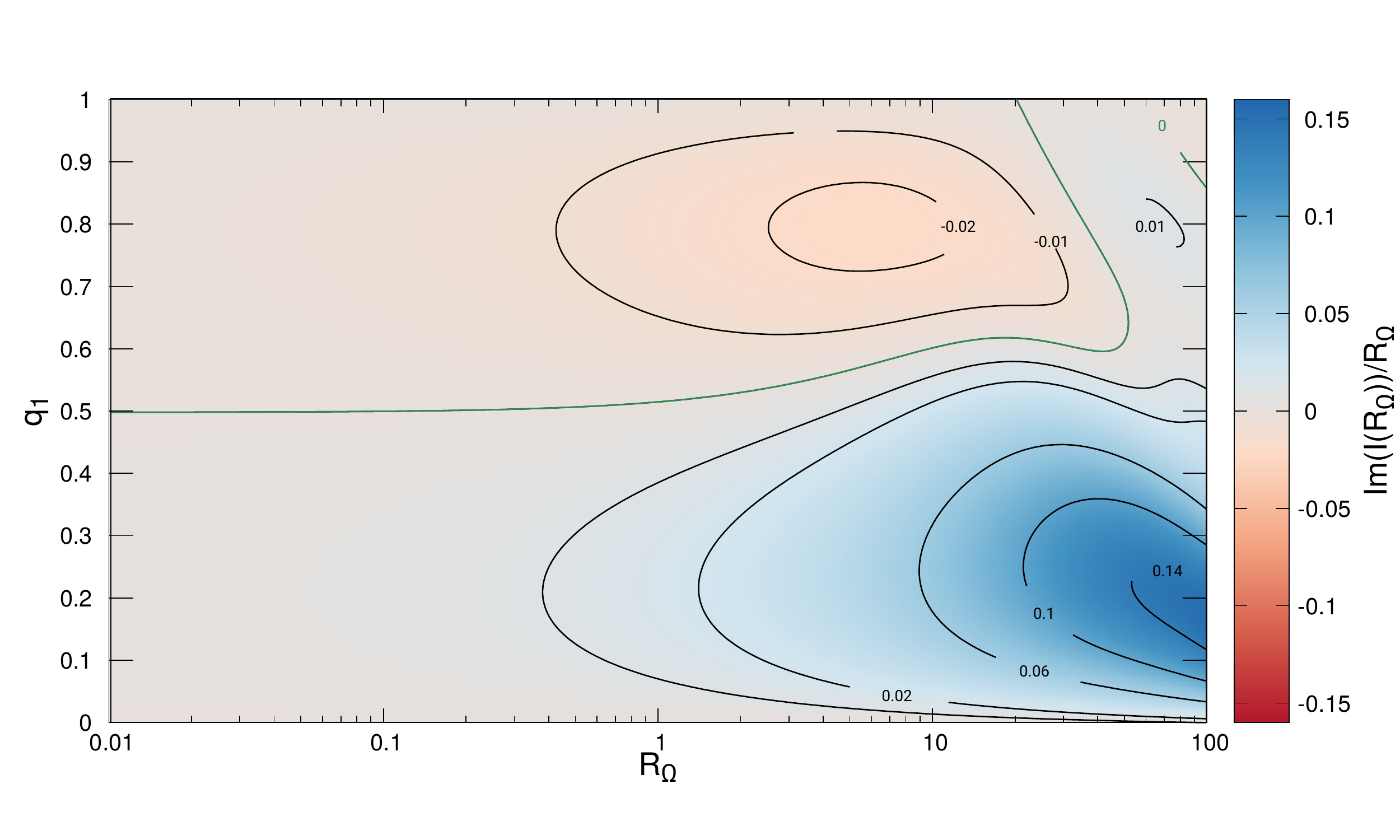}
    \end{minipage}
    \caption{Contour levels of $\Re(I)/\rv^2$ (top panel) and $\Im(I)/\rv$ in the plane $(\rv,q_1)$ where $q_1$ is the (normalized) mass of the first binary component. A value of $\tkmin=0.3$ is assumed for the computation of $I(\rv)$. The green contours indicate the zero level. The increase in the magnitude of $\Re(I)$ towards smaller $\rv$ is artificially caused by the normalization $\Re(I)/\rv^2$ adopted here. $\Re(I)$ linearly depends on $\rv$ for $\rv\lesssim 1$, but exhibits a quadratic dependence $\propto\rv^2$ for $\rv\gg 1$. Both $\Re(I)$ and $\Im(I)$ can be positive or negative depending on $(\rv,q_1)$.}
    \label{fig:fdmBin}
\end{figure*}

The physical origin of this behavior stems from the wave and diffusive nature of the free Schr\"odinger equation, in which $\beta=\hbar/2m_a$ can also be regarded as a diffusion coefficient \cite[see][]{nelson:1966,mita:2021}. Diffusion (of the condensate wave function) in 3-dimensional space then implies $\langle r^2\rangle = 6\beta t$. Therefore, the timescale corresponding to a diffusion length $\sim \pi/k_\text{kmin}$ is $t_\text{diff}=(\pi/k_\text{min})^2/6\beta$ or, equivalently,
\begin{equation}
    \label{eq:tdiff}
    \frac{{\tilde t}_\text{diff}}{2\pi} = \bigg(\frac{\pi}{12}\bigg)\frac{\rv}{\tkmin^2}\;.
\end{equation}
In Fig.~\ref{fig:transM0}, this characteristic time is indicated as vertical arrows. Eq.~(\ref{eq:tdiff}) reasonably captures the $\rv$ and $\tkmin$ dependence of the timescale marking the transition from a (radially) diffusive regime to damped oscillations around the steady-state result.

\subsection{Compact circular binary}

The same techniques can be used to derive analytic results in the compact binary case. Let $M$ denote the total mass of the binary system and $q_1 M$ and $q_2 M$ (with $q_1+q_2=1$) the mass of the individual components. To account for their different distance to the binary center-of-mass, we must also change the argument of the spherical Bessel function in Eq.~(\ref{eq:scat}). As a result, the denominator of Eq.~(\ref{eq:Im}) is replaced by
\begin{multline*}
    q_a^2\left[\sllb{a}{a}(m,\rv,t)-S_{l,l-1}^{a,a\ *}(m+1,\rv,t)\right]
        \\+(-1)^m q_a q_b \left[\sllb{a}{b}(m,\rv,t)+S_{l,l-1}^{a,b\ *}(m+1,\rv,t) \right]\;.
\end{multline*}
The steady-state solution reads
\begin{widetext}
\begin{align}
    \sllb{a}{b}&(m,\rv)= \\
    &
    \frac{\pi \rv}{4}\left \lbrace \begin{array}{lc}
       \frac{i}{m} \left[\hone{l}{q_a\sqrt{ m \rv}}\bessJ{l-1}{q_b \sqrt{m \rv}}-\frac{i\ \mathrm{K}_{l+1/2}(q_a\sqrt{ m \rv})\mathrm{I}_{l-1/2}(q_b\sqrt{ m \rv}) }{\sqrt{q_a q_b  m \rv}} +\frac{q_b^{l-1}}{q_a^{l+1}}\frac{2i}{|m|\rv}\right]  & (q_a>q_b) \wedge m\neq 0 \\
        \rv (\frac{q_b}{q_a})^{l-1}\ \frac{4(3+4 l (2+l))q_a^4 +2(9-4l^2)q_a^2 q_b^2+(3+4l(l-2))q_b^4 }{(9-40l^2+16l^4)q_a^2} & (q_a>q_b) \wedge m=0\\
        \frac{i}{m} \left[\bessJ{l}{q_a\sqrt{ m \rv}}\hone{l-1}{q_b\sqrt{ m \rv}}+\frac{i\ \mathrm{I}_{l+1/2}(q_a \sqrt{ m\rv})\mathrm{K}_{l-1/2}(q_b\sqrt{ m \rv}) }{\sqrt{q_a q_b  m \rv}}\right] & (q_a<q_b) \wedge m\neq 0  \\
        -\rv  (\frac{q_a}{q_b})^l\ \frac{(2l-3)q_a^2-(2l+3)q_b^2}{9-40l^2+16l^4} &(q_a<q_b) \wedge m=0
    \end{array}\right. \nonumber 
\end{align}
\end{widetext}
with $(q_a,q_b)$ equal to $(q_2,q_1)$ (resp. $(q_1,q_2)$) if the DF acting on object 1 (resp. object 2) is desired. Like in the single perturber case, this expression is not valid for $l=1$ and $m=0$ and a lower cut-off $\tkmin$ must be introduced to regularize the infrared divergence. An analytic solution can still be found in this case, but it is lengthy and, therefore, presented in Appendix~\ref{sec:sllbl1} solely. 

The resulting function $I(\rv)$ is shown in Fig.~\ref{fig:fdmBin} assuming $\tkmin=0.3$ for illustration. The top and bottom panels display contour levels of the real and imaginary part $\Re(I)$ and $\Im(I)$ as a function of $\rv$ and the mass $0<q_1<1$ of the first binary component. There are combinations of $\rv$ and $q_1$ for which the radial and tangential component of DF change sign. 
When $q_1>0.5$, the radial DF force is mostly directed inward ($\Re(I)>0$) while the tangential component points forward along the direction of motion ($\Im(I)<0$) for $0.5<\rv<20$. When $q_1<0.5$, DF pushes the point mass outward ($\Re(I)<0$) for $\rv<0.2$ while it generally slows it down in the tangential direction ($\Im(I)>0$).

\begin{figure*}
    \centering
    \setlength\abovecaptionskip{-3.0\baselineskip}
    \input{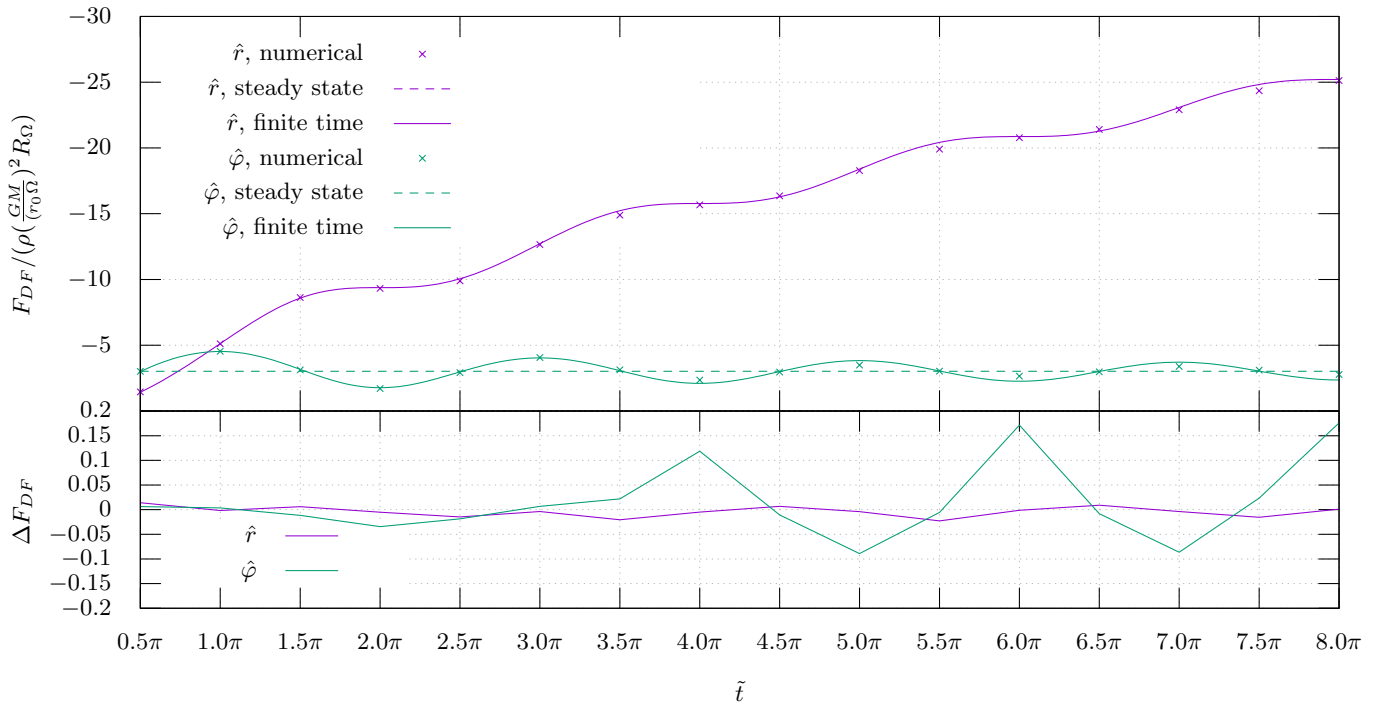}
    \caption{{\it Top panel}~: radial ($\hvr$) and tangential ($\hpv$) components of the DF force in the finite time perturbation case for our fiducial single perturber. The DF force is normalized to  $\rho\rv\left(\frac{G M}{\Omega r_0}\right)^2$. The cross symbols are the outcome of a numerical implementation of the Lippmann-Schwinger approach on a 3-dimensional grid, whereas the solid and dashed curves show the analytic predictions obtained from the Madelung approach (see text for details). We have adopted a cut-off wavenumber $\tkmin=1/8$ which matches the size of our simulation boxes.
    The steady-state tangential DF is shown as the horizontal dashed line. In the radial direction, the steady-state DF is $\simeq -76.6$ for $\tkmin=1/8$. Hence, it is not visible on the figure. 
    {\it Bottom panel}~: fractional difference between our wave and hydrodynamic predictions of the radial and tangential DF.}
    \label{fig:lsMComp}
\end{figure*}

\subsection{Comparison between Lippmann-Schwinger and Madelung DF}
\label{sec:NumCom}

As discussed above, both treatments have their advantages. While the Lippmann-Schwinger approach can easily incorporate wave superposition and interference effects, the Madelung approach allows for the fully analytical solution to DF presented in Sec.~\ref{sec:sty} owing to its convenient hydrodynamic form. More specifically, wave interference in the atmosphere of virialized halos creates a haze of fluctuating granules or wave packets with a distribution of velocities \cite{schive/etal:2014a,schive/etal:2014b,schwabe/etal:2016}. Such an interference pattern can be easily accounted for in the Lippmann-Schwinger formulation. However, this inclusion is less straightforward in the Madelung approach, and our implementation of the latter assumes zero velocity dispersion. Therefore, it is instructive to compare the DF obtained in both implementations.

To compute DF in the Lippmann-Schwinger approach, we calculate the overdensity $\alpha$ directly from Eqs.~(\ref{eq:delPSi}) and~(\ref{eq:alphaQMRaw}) over the simulation box before integrating Eq.~(\ref{eq:dynFricRaw}) numerically. Namely, on rewriting Eq.~(\ref{eq:dynFricRaw}) in terms of dimensionless variables, we arrive at
\begin{equation}
    \mathbf F_{\mathrm{DF}}(\tilde t)= \left(\frac{G M}{\Omega r_0}\right)^2\rho\rv \int \mathrm{d}^3\tilde u \left(\frac{\tvu}{\tilde u^3}\right) \frac{\alpha(\tvu,\tilde t)}{\alpha_0}\;.
    \label{eq:FdfNum}
\end{equation}
where $\alpha_0$ is the characteristic amplitude of the overdensity wake, Eq.~(\ref{eq:alpha0}). The field $\alpha(\tvu,\tilde t)$ is then computed on a regular cubical of size $8r_0$ with $64^3$ mesh points (see Sec.~\S\ref{sec:numerics} for the details of the numerical implementation). Splines are used to interpolate this data and numerically evaluate Eq.~(\ref{eq:FdfNum}). 

The resulting DF is shown in the top panel of Fig.~\ref{fig:lsMComp} as the cross symbols for our fiducial single perturber (i.e. $\rv=4$ and $\rs=0.1$ ) turned on at $t=0$. It is compared to the steady state solution Eqs.~(\ref{eq:slls}) -- (\ref{eq:styM0}) obtained from the Madelung approach (dashed curves) and to the finite perturbation time result Eq.~\ref{eq:sllt} (solid curves) assuming a lower cut-off $\tkmin=1/8$. This wavenumber value corresponds to a scale equal to the side length $8r_0$ of the cubical box. This choice is motivated by the absence of power on scales larger than the simulation box. 
The bottom panel of Fig.~\ref{fig:lsMComp} displays the fractional difference between the numerical prediction of the Lippmann-Schwinger approach and the analytic prediction of the Madelung formulation. 

As expected, the radial DF strongly depends on the behavior of the purely real amplitude $S^\text{Ftp}_{1,0}$. At early times, $S^\text{Ftp}_{1,0}$ is not very sensitive to the exact choice of $\tkmin\sim 0.1$ (see Fig.~\ref{fig:transM0}). Notwithstanding, a low cut-off wavenumber implies that the radial DF grows through diffusion for a longer time before oscillating around the steady-state solution. For $\tkmin=1/8$, the steady-state radial DF is $\approx -76.6$ in the scales of the figure and, therefore, does not appear on it. As a result, the radial, steady-state solution is a poor approximation to the finite time perturbation result so long as $\tilde t\lesssim$ $O(10-100)$ rotations. By contrast, the tangential part rapidly initiates damped oscillation around the steady state solution. One should however bear in mind that neither the radial nor the tangential DF attain the equilibrium steady-state regime in a finite time.

The finite time perturbation results agree very well for both the tangential and radial DF component. Although our Lippmann-Schwinger implementation of DF is more complete from a modeling point of view (it includes the granularity of the FDM background), it is matched by the Madelung analytical prediction to a high degree.
In the radial direction, the relative difference between the wave-like and fluid-like predictions never exceeds 3\%. 
In the tangential direction, the agreement is equally good during the first rotation, yet the relative difference between the wave and hydrodynamic prediction increases with time. This reflects the fact that, in the Lippmann-Schwinger approach, the tangential component converges significantly faster to steady-state.

We emphasize that the comparison carried out here is somewhat limited since the regime $\ms\gg 1$ (our fiducial single perturber) solely was tested. A more exhaustive comparison is beyond the scope of this paper. Notwithstanding, the differences in the $\alpha$-predictions seen in the binary case with $\ms\lesssim 1$ (Fig.~\ref{fig:alphaQMR01}) suggest that the agreement between the Lippmann-Schwinger and Madelung DF shall worsen in the single perturber case as the FDM velocity dispersion $\sigma$ or, equivalently, the Mach number $\ms$ decreases. While a non-vanishing $\sigma$ likely changes the convergence rate to steady-state (i.e. the decay timescale of the oscillatory envelope), it is unclear whether the steady-state solution is also affected.

\begin{table*}
    \centering
    \begin{tabular}{c|c|c|c|c|c|c|c|c|c}
         & & & \multicolumn{2}{c|}{CDM}& \multicolumn{2}{c|}{FDM linear}& \multicolumn{3}{c}{FDM circular} \\
         \hline
         ~n~ & $r_0$ $[\Kpc]$ & $M$ $[10^{5}M_\odot]$ & $C$ & $\tau$ $[\mathrm{Gyr}]$ & $C$ & $\tau$ $[\mathrm{Gyr}]$ & $\rv$ & $\Im(I)$ & $\tau$ $[\mathrm{Gyr}]$ \\
         \hline
         1 & 7.60 & 0.37 & 4.29 & 112 & 2.46& 215 & 17.8 & 1.46 &362 \\
         2 & 1.05 & 1.82 &3.32 & 9.7 & 1.88 & 12 & 10.08 & 1.64 & 14 \\
         3 & 0.43 & 3.63 & 2.45 & 0.62 & 0.29 & 2.2 & 1.94 & 0.39 & 1.63 \\
         4 & 0.24 & 1.32 & 2.50 & 0.37 & 0.033 & 10 & 0.62 & 0.078 & 4.23 \\
         5 & 7.79 & 1.76 & 3.46 & 21.3 & 2.32 & 31 & 15.58 & 1.41 & 51
    \end{tabular}
    \caption{A comparison between the orbital decay timescale obtained for a perturber moving linearly in a CDM medium ("CDM"), in a FDM background ("FDM linear"), and for a circularly-moving perturber in FDM ("FDM circular"). Results are shown for the 5 globular clusters of the Fornax dwarf spheroidal. This table is adapted from Table I of \cite{hui/etal:2017} from which the "CDM" and "FDM linear" predictions are taken.}
    \label{tab:orbdecay}
\end{table*}

\section{Discussion}
\label{sec:discussion}

In this Section, we present two astrophysical applications of our results. We also speculate on the behavior of DF for self-interacting axions.

\subsection{Orbital decay of globular clusters}

Massive satellites orbiting around a galaxy are affected by the dynamical friction caused by the dark matter halo. This leads to their orbital decay if the dark matter density is high enough and its velocity dispersion is low \cite{chandrasekhar:1943}.
Following \cite{hui/etal:2017}, let $L=M\Omega r_0^2$ be the angular momentum of, say, a globular cluster of mass $M$. The corresponding DF timescale is given by
\begin{equation*}
    \tau=\frac{L}{r_0 |\hpv\cdot\fdf|}\;,
\end{equation*}
where $r_0 |\hpv\cdot\fdf|$ is the torque produced by DF. For the slow orbital decay of circular orbits, this timescale can be expressed as
\begin{equation}
    \label{eq:orbdecay}
    \tau=\frac{M(<r_0)^{3/2}}{4\pi G^{1/2} M \rho r_0^{3/2} \Im(I)}\;,
\end{equation}
with $M(<r_0)$ being the total (dark matter and baryons) mass enclosed within the orbital radius.

A well-known example of application is the Fornax dwarf spheroidal, which contains 5 widely spread globular clusters. 
These should have already spiraled toward the center and merged with the nucleus \cite{Tremaine1976}. To resolve this issue, various explanations have been put forward, including the presence of a constant-density core that can stall the inspiral \cite{Read_2006}. However, this might not be sufficient to explain the Fornax observations \cite{Meadows2020,BarNitsanBlas2021}. Ref.~\cite{hui/etal:2017} assessed how the DF timescale is modified if the dark matter in Fornax and, more generally, in dwarf galaxies is comprised of FDM rather than CDM. For this purpose, they derived an expression for the DF produced by a FDM medium on a perturber in linear motion (see their Sec.~\S~III.J).

To compare our circular motion prediction with their result, we use the fact that their drag coefficient $\mathcal{C}$ corresponds to our $\Im(I)$ (the radial DF is irrelevant here) and, moreover, their choice of $k r$ is equivalent to our $\rv/2$. Assuming likewise an axion mass $m_{18}=3\times 10^{-4}$, the nonlinearity scale is $\lambda_\text{NL}\sim 10^{-3}\pc$ much smaller than the orbital radius of the globular clusters ($r_0\approx 1\Kpc$) considered here. Therefore, our linear response theory can be safely applied here. Furthermore, our analysis of the finite time perturbation indicates that our steady-state solution provides a reasonable approximation to the tangential DF at all time (see Fig.~\ref{fig:lsMComp}). Therefore, we shall use it to compute the orbital decay timescales determined by Eq.~(\ref{eq:orbdecay}).

To illustrate how the DF decay timescale changes quantitatively when our circular motion result is used instead of the linear motion expression of \cite{hui/etal:2017}, we have extended their Table.I into a new table \ref{tab:orbdecay}, in which we quote $\tau$ obtained from our "FDM circular" and their "FDM linear", respectively. Using our circular motion DF increases or decreases $\tau$ by up to $\sim 70$\% depending on the value of $\rv$. 
To understand this, recall that the (tangential) FDM friction coefficient $\Im(I)$ peaks for values of $\rv\sim \mathcal{O}(1)$ (see Fig.~\ref{fig:fdmGasComp}). This corresponds to a configuration in which the perturber's motion is best synchronized with the gravitational wake it generates, thereby increasing the strength of the tangential DF. 
At large $\rv\gg 1$, the weaker synchronization between the perturber's motion and the induced density wake reduces the tangential DF. Overall, there is no dramatic change from the linear motion calculation.

\subsection{Stagnation of binary inspiral}

Dynamical friction generally leads to the dissipation of energy and angular momentum. We will now demonstrate that, in a FDM background, DF can stall the orbital evolution. In what follows, $M$ is the total binary mass and $q_1M$, $q_2M$ are the masses of the components. 

Consider first the motion of the binary center-of-mass of position $\vrcm$. If the binary is not of equal mass, a net force accelerates the center-of-mass according to
\begin{align}
    M \frac{\mathrm{d}^2\vrcm}{\mathrm{d}t^2}&=\vvf_\text{DF,1}+\vvf_\text{DF,2}\\
    &= -4\pi\rho\left(\frac{GM}{\Omega r_0}\right)^2
    \Big[\big(\Re(I_1)-\Re(I_2)\big)\,\hvr(t) \nonumber \\
    &\qquad +\big(\Im(I_1)-\Im(I_2)\big)\,\hpv(t) \Big]\nonumber \;,
\end{align}
where $\hvr(t)$ and $\hpv(t)$ are unit vectors in the $x-y$ plane directed along the component separation vector $\vr(t)\equiv\vr_2(t)-\vr_1(t)$ and perpendicular to it. 
Decomposing the total force into this radial and tangential direction,
\begin{align}
    F_r&\equiv4\pi \rho\left(\frac{GM}{\Omega r_0}\right)^2\big[\Re(I_1)-\Re(I_2)\big]\\
    F_\varphi&\equiv4\pi \rho\left(\frac{GM}{\Omega r_0}\right)^2
    \big[\Im(I_1)-\Im(I_2)\big] \nonumber \;,
\end{align}
and rewriting the equation of motion in Cartesian $x-y$ coordinates (we omit the $z$ component as it is irrelevant), we get 
\begin{equation}
    M\frac{\mathrm{d}^2\vr_{\text{CM}}}{\mathrm{d}t^2}=-
    \begin{pmatrix}F_r\cos(\Omega t)-F_\varphi\sin(\Omega t)\\
                   F_r \sin(\Omega t)+F_\varphi\cos(\Omega t)
    \end{pmatrix}\;
\end{equation}
Assuming the steady state solution for $I_1$ and $I_2$ makes $F_r$ and $F_\varphi$ independent of time and the differential equation straightforward to solve.
For a binary system initially at rest at $\vrcm(0)=(F_r/M\Omega^2,F_\varphi/M\Omega^2)^\top$, the motion of the center of mass describes a circle about the origin $\vr=0$ at a frequency $\Omega$. The radius of this circular orbit is 
\begin{align}
    r_\text{DF} &= \frac{1}{M\Omega^2}\sqrt{F_r^2+F_\varphi^2} \\
    &\simeq 2.5\times 10^{-28}\pc \left(\frac{\rho}{\mpppc}\right) \nonumber \\
    &\qquad \times \left(\frac{M}{M_\odot}\right)\left(\frac{\Omega}{\yr^{-1}}\right)^{-4}\left(\frac{r_0}{\pc}\right)^{-2} \nonumber \\
    &\qquad\times \sqrt{\big(\Re(I_1)-\Re(I_2)\big)^2+\big(\Im(I_1)-\Im(I_2)\big)^2}
    \nonumber \;.
\end{align}
Therefore, $r_\text{DF}$ reassuringly is orders of magnitude smaller than the orbital radius even for $\tkmin\ll 1$. More precisely, parameter values as extreme as $\tkmin=0.001$ and $\rv=100$ are required such that the difference between $I_{1}$ and $I_{2}$ reaches $\mathcal{O}(10^5)$. Even in this case would the radius $r_\text{DF}$ be of the order of $10^{-23}$ solely, that is, orders of magnitude below $\lambda_\text{NL}$. There seems to be no realistic scenario in which the motion of the center-of-mass caused by DF can have any significant impact on the motion of the perturber.

Turning now to the center-of-mass frame, the energy $E=\frac{1}{2}\mu \dot\vr^2-\frac{GM\mu}{r}$ and angular momentum $\vvl=\mu\vr\times\dot\vr$ of the binary, where $\mu=q_1 q_2 M$ is the reduced mass, evolve according to
\begin{align}
    \frac{dE}{dt} &= \dot\vr\cdot\left(q_1 \vvf_\text{DF,2} - q_2\vvf_\text{DF,1}\right) \\
    \frac{d\vvl}{dt} &= \vr \times \left( q_1\vvf_\text{DF,2} - q_2 \vvf_\text{DF,1}\right) \nonumber \;.
\end{align}
For the homogeneous medium and circular motions considered here, we have $\vvl = L \hvz$ with
\begin{equation}
    \frac{dL}{dt} = -4\pi r_0\rho\left(\frac{GM}{\Omega r_0}\right)^2
    \big(q_1 \Im(I_2) + q_2\Im(I_1)\big) \;.
\end{equation}
Assuming an adiabatic sequence of circular orbits, we have $L^2 = GM^2\mu r_0$ and, therefore,
\begin{equation}
    \label{eq:dr0dt}
    \frac{dr_0}{dt} = -8\pi\rho\, \sqrt{\frac{G r_0^5}{\mu}}\,
    \big(q_1 \Im(I_2) + q_2\Im(I_1)\big) \;.
\end{equation}
The radial part of DF is irrelevant here as it mainly affects the eccentricity \cite{burns:1976,kim/etal:2007}. Note also that the right-hand side of Eq.~(\ref{eq:dr0dt}) is invariant under the exchange of indices $1\leftrightarrow 2$.
Since the DF acting on the binary depends on $r_0$, $m_a$ and $M$ through the parameter $\rv$, it is convenient to introduce a characteristic orbital radius $r_\Omega$. For simplicity, we set
\begin{align}
    \label{eq:romega}
    r_\Omega &\equiv\frac{1}{GM} \left(\frac{\hbar}{2 m_a}\right)^2 \\
             &\simeq 171.2\pc\, m_{18}^{-2}\left(\frac{M}{M_\odot}\right)^{-1}\nonumber
\end{align}
which corresponds to the orbital radius such that $\rv=1$.

\begin{figure}
    \centering
    \hspace*{-0.8cm}    
    \setlength\abovecaptionskip{-0.7\baselineskip}
    \input{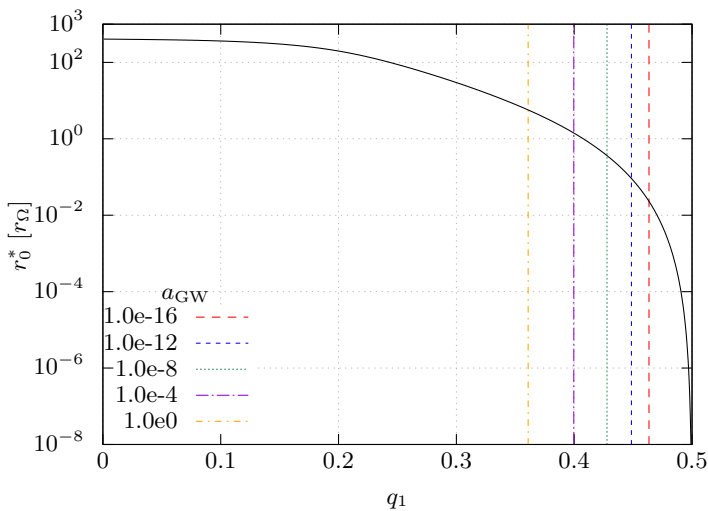}
    \caption{Stable fixed points $r_0^*$ of the one-dimensional dynamical system Eq.~(\ref{eq:dr0dtGw}) as a function of $q_1$ ($q_2=1-q_1$). When $\agw=0$, there is one stable orbital radius for each $q_1$ shown as the solid (black) curve. When $\agw\neq0$, there are no stable orbits for $q_1$ in the range $[q_\text{min},0.5]$. Vertical lines mark $q_\text{min}$ for the different values of $\agw$ indicated on the figure. For each $q_1$ in the range $[0,q_\text{min}[$, there is, again, a unique stable orbit. Whenever they exist, the stable orbital radii $r_0^*$ change by at most 1\% as $\agw$ is varied in the range $0<\agw<1$.  Therefore, the solid (black) curve computed for $\agw=0$ accurately characterize the fixed point also for $0<\agw<1$ (leftward of the corresponding vertical line).}
    \label{fig:fixedPoint}
\end{figure}

Furthermore, we take into account the energy/angular momentum loss by radiation of gravitational waves (GW). The orbit average change of $r_0$ reads (\cite{Peters64}) 
\begin{equation*}
    \left< \frac{dr_0}{dt}\right>=-\frac{64}{5} \frac{G^3 \mu M^2}{c^5 r_0^3}\;.
\end{equation*}
Adding this loss term to Eq.~(\ref{eq:dr0dt}) yields
\begin{align}
    \label{eq:dr0dtGw}
    \frac{dr_0}{dt} &=-8\pi\rho \sqrt{\frac{G r_\Omega^5}{\mu}} 
    \Bigg[\frac{8}{5 \pi} \agw \left(\frac{r_0}{r_\Omega}\right)^{-3} \\
    &\qquad +
    \left(\frac{r_0}{r_\Omega}\right)^{5/2}\big(q_1 \Im(I_2) + q_2\Im(I_1)\big)\Bigg]
    \nonumber \;.
\end{align}
The relative strength of the loss by gravitational waves is given by
\begin{align}
    \agw&\equiv\sqrt{\frac{G^{5}\mu^{3}M^4}{c^{10}r_\Omega^{11}\rho^2 }}\\
    &\simeq5.03\times 10^{-34}\left(\frac{\mu}{M_\odot}\right)^{3/2}\left(\frac{M}{M_\odot}\right)^2\left(\frac{r_\Omega}{\pc}\right)^{-11/2} \nonumber \\
    &\qquad \times\left(\frac{\rho}{\mpppc}\right)^{-1} \nonumber \;.
\end{align}
For our fiducial binary system (see~\ref{eq:BinExample}) and a FDM density of $\rho=0.01\mpppc$ comparable to that of the solar neighborhood, this gives
\begin{align*}
    r_\Omega&\simeq8.56\pc \\
    \agw&\simeq1.67\times10^{-33}\;.
\end{align*}
For a given choice of $q_1$ and $q_2$, the friction coefficient $ \Im(I_{1/2}) $ can change sign as $\rv$ is varied (seen in Fig.~\ref{fig:fdmBin}). Friction becomes a "thrust" (rather than a "drag") when the binary extracts angular momentum from the FDM medium. Furthermore, at fixed $m_a$ and $M$, the value of $\rv$ depends only on $r_0$ (through Kepler's third law). Therefore, the right-hand side of Eq.~(\ref{eq:dr0dtGw}) should be regarded as a function of $r_0$ only, say, $g(r_0)$. This one-dimensional dynamical system can exhibit stable fixed points for $r_0\equiv r_0^*$ whenever $g(r_0^*)=0$ and $g'(r_0^*)<0$. 

When there is no loss from GW emission, there is exactly one stable orbit for each choice of $q_1$ except $q_1=0.5$. In Fig.~\ref{fig:fixedPoint}, the solid (black) curve shows the stable orbital radius $r_0^*$ as a function of $q_1$ for our fiducial binary system. 
At small $q_1$, the stable radius is $r_0^*\approx400$ (in unit of $r_\Omega$) and varies slowly until $q_1\approx 0.2$, beyond which the stable orbit quickly drops towards smaller radii. Adding the GW emission affects the stable orbital radius only marginally at low $q_1$: the change never exceeds 1\% even for $\agw=1.0$. The most notable effect of a non-vanishing $\agw$ is to prevent stable fixed points in the $q_1$-range $[q_\text{min},0.5]$. In Fig.~\ref{fig:fixedPoint}, vertical lines mark $q_\text{min}$ for different values of $\agw$. The $q_1$-range for which there are no stable orbits grows from 0.5 downward with increasing values of $\agw$. This follows from the $r_0^{-3}$ dependence of the GW loss, which becomes relevant only when the stable orbit wanders to the small radii obtained for $q_1\lesssim 0.5$. A larger $\agw$ increases the orbital radius below which GW emission dominates over DF. Note that, since the DF only depends on $m_a$ via $\rv$ solely, any change in the axion mass can be absorbed by a rescaling of $r_\Omega$ according to Eq.~(\ref{eq:romega}). In other words, changing $m_a$ leaves Fig.~\ref{fig:fixedPoint} unchanged.

For our fiducial choice of parameters, the characteristic radius $r_\Omega\sim 10\pc$ is about 3 orders of magnitude larger than any viable upper limit on galactic binary separations \cite{Oepik:1924,DucheneKraus:2013}. Consequently, binaries of mass $M\sim \mathcal{O}(10)\msun$ and viable separations will not inspiral through the fixed point unless $q_1$ is very close to 0.5. However, as the axion mass is increased, the characteristic radius drops according to $r_\Omega\propto m_a^{-2}$ so that stable orbits with realistic $r_0^*$ appear for a larger range of $q_1$. In the absence of external perturbations \\cite[see, e.g.,][for a discussion]{ginat/perets:2021}, binaries would eventually stagnate around this stable orbit (provided they started with a larger orbital radius). This could have an impact on merger rates.

\subsection{Axion self-interactions}

Axions self-interactions will change the structure of the Green's function. For axions which acquired a mass through non-perturbative effects (such as the QCD axion), this self-interaction is attractive and, owing to the enormous phase space density, can counteract the quantum pressure. This leads to an instability which has been explored in, e.g.,  \cite{vv1973,Riotto:2000kh,chavanis:2011,chavanis/delfini:2011,eby/suranyi/etal:2015,chavanis:2016,levkov/etal:2017,eby/ma/etal:2017,helfer/marsh/etal:2017,Desjacques:2017fmf,glennon/prescodweinstein:2021,glennon/etal:2022}. 

At the level of the Green's function of the linearized theory, a self-interaction manifests itself as a pressure with an (effective) sound speed $c_s^2>0$ or ($c_s^2<0$) if the self-interaction is repulsive (attractive). In plain words, $G$ is of the form
\begin{equation}
\tilde G(\vk,\omega) = \bigg(c_s^2 k^2 + \frac{\hbar^2k^4}{4 m_a^2} - \omega^2\bigg)^{-1}
\end{equation}
in Fourier space.
Our present work, together with the analysis of \cite{DesNuBu21}, suggests that the resulting DF should be free of any infrared and ultraviolet divergence since, for $\omega=0$, the divergence of $G$ in the limit $k\to 0$ is only quadratic whereas its high-$k$ limit is regularized by the $k^4$ term. We defer a thorough exploration of this case to future work.

\section{Conclusion}
\label{sec:conclusion}

We investigated the dynamical friction (DF) acting on circularly moving perturbers in a background of FDM particles. 
Starting from the Gross-Pitaievskii-Poisson system which describes a self-gravitating FDM medium, we considered two different routes to solve for the density wake and DF: the Madelung (hydrodynamic) and Lippmann-Schwinger (wave) approach.
Although the latter can more straightforwardly account for the fluctuating nature of FDM halo atmospheres, the former is more amenable to an analytic treatment of DF. For this reason, our hydrodynamic implementation assumes a perfectly uniform FDM background, whereas our wave implementation describes the medium as a superposition of FDM quasi-particles. Furthermore, we restricted our analysis to linear response theory. The astrophysical systems considered here are well within the validity range of this linear approximation. 

We derived a fully analytical solution to the dynamical friction using the Madelung formulation. Our circular-orbit solution, based on the approach outlined in \cite{DesNuBu21}, covers steady-state as well as the finite time perturbation case (the perturber is turned on at $t=0$). Although it does not include the velocity dispersion of FDM quasi-particles, it provides a versatile tool to explore DF for a wide range of parameters.
We compared the two approaches at the level of the density wake and the DF produced by single and binary compact perturbers in circular motion. The velocity dispersion $\sigma$ of FDM quasi-particles generally lowers the overall density contrast, the effect increasing with smaller values of the Mach number $\ms=v_\text{circ}/\sigma$.
Moreover, for the finite time perturbation, our analytical solution to DF agrees very well with that extracted from our limited numerical investigations of the wave formulation.

The distinctive form of the FDM and gaseous Green's functions considered here and in \cite{DesNuBu21}, respectively, leads to critical differences in the behavior of the dynamical friction. While the ultraviolet divergence (seen for supersonic motion in the gaseous medium) is no longer present in the FDM case, the latter exhibits instead an infrared divergence which originates from the (also) diffusive nature of the free Schr\"odinger equation. Our analysis of the finite time perturbation case reveals that the density wake produced by the perturber(s) diffuses through the medium with a diffusion coefficient $\beta=\hbar/2m_a$ ($m_a$ is the axion mass). Only when the characteristic diffusion length reaches the size of the system does DF stabilize around the steady-state result. This diffusive process affects the radial component of DF solely. Once the initial perturbation has diffused through the whole medium, both the radial and tangential DF oscillate about the steady-state solution with a decaying envelope. Strictly speaking, steady-state is thus never attained within a finite time. Our numerical implementation of the wave approach, which includes a non-vanishing FDM velocity dispersion, indicates that the convergence rate is somewhat sensitive to the value of $\sigma$. Although we have not determined the extent to which the steady-state solution depends on $\sigma\ne 0$, we speculate (in light of our single perturber test case) that the Madelung prediction remains valid so long as $\ms\gg 1$. Notice also that the damped oscillations seen in the tangential DF only arise from multipoles with $(l,m)\ne(1,0)$, which are insensitive to the outer boundary conditions. These oscillations thus have a physical origin different from those studied in \cite{vicente/etal:2019}.

We applied our results to two different astrophysical scenarios. Firstly, we revisited the DF decay timescale of the 5 Fornax globular clusters using our circular-motion prediction, and compared them to the linear-motion estimates of \cite{hui/etal:2017}. In the circular setting, the imaginary part $\Im(I)$ of the "complex friction" $I$ encodes the drag in the direction parallel to the instantaneous perturber's velocity. At fixed axion mass, our circular-motion result increases the decay time relative to the linear-motion prediction when $\rv\propto \sqrt{r_0}\gg 1$, i.e. for large orbital radii, and decreases it for small $r_0$. This follows from the fact the interaction between the perturber and its wake is maximized for $\rv\sim \mathcal{O}(1)$. While the relative change can be as large as 70\%, our revised DF decay timescales are still larger than that obtained for a CDM medium. 
Secondly, we explored the stagnation of compact binary inspirals assuming an adiabatic sequence of circular orbits. The motivation is the presence of a novel feature in the tangential DF acting on binaries: it can change sign and act as a "thrust" (rather than a "drag") for some range of mass ratios. This effect could stall binary inspirals and lead to their stagnation if there are stable orbits (provided that external perturbations are small). This might also impact binary merger rates. 

Ignoring energy/angular momentum loss through gravitational wave (GW) emission, a stable circular orbit exists for any compact binary, except for equal-mass systems. The inclusion of GW emission prevents the existence of stable orbits for near equal-mass binaries. A stronger GW loss implies a smaller range of stable circular orbits but, for conservative assumptions, this effect is small. For our fiducial axion mass $m_a\sim 10^{-18}\eV$, most of the stable orbits are far outside the range of viable binary orbital radii. However, larger axion masses would move them in the interesting range $r_0\lesssim 0.01\pc$.
To conclude, we stress that this effect is not restricted to FDM backgrounds. It can arise in any medium as long as the tangential DF can change sign, and sometimes be a thrust rather than a drag. Note also that it is physically different from the core-stalling discussed in \cite{kaur/etal:2018}. It would be interesting to explore this further taking into account the orbital eccentricity.

\acknowledgments

We thank Adi Nusser for useful discussions and Barry Ginat, Rodrigo Vicente for helpful comments on an earlier version of the manuscript.
We acknowledge support by the Israel Science Foundation (ISF) grants no 2562/20.
This work was performed in part at Aspen Center for Physics, which is supported by National Science Foundation grant PHY-1607611. VD would also like to thank Fabian Schmidt and the Max-Planck Institute of Astrophysics (Garching) for hospitality during the final redactional stage of this paper. 

\bibliographystyle{prsty}
\bibliography{bibliography}

\onecolumngrid

\appendix

\section{Born approximation to the perturbed wave function}
\label{sec:LippSchw}

In the (first order) Born approximation, the Lippmann-Schwinger approach returns
\begin{equation}
    \delta \psi(\vr,t)=m_a\int \mathrm{d}r'^3\int \mathrm{d}t' G_\text{ret}(\vr-\vr',t-t')\, \Phi(\vr',t')\,\psi_0(\vr',t')\;,
\end{equation}
where the building blocks of this expression are given by Eqs.~\ref{eq:psi0}, \ref{eq:GreensQM} and \ref{eq:Phi}.
The Convolution theorem can be apply for the computation of this integral. This requires the combined Fourier transform of $\Phi$ and $\psi_0$:
\begin{align}
     \tilde{(\Phi\psi_0)}(\vk,\omega)&=-\int \mathrm d r^3 \int \mathrm d t\ e^{i \omega t -i \vk\cdot\vr}\,\Phi(\vr,t)\,\psi(\vr,t)\\
    &=-\int \mathrm d r^3 \int \mathrm d t\ e^{i \omega t -i \vk\cdot\vr} h(t)\,\frac{G M}{|\vr -\vr_p(t)|}\,\sqrt{\rho}\, e^{-i \omega_0 t+i \vk_0 \cdot\vr}\nonumber\\
    &=-G M \sqrt \rho \int \mathrm d t\ e^{i (\omega-\omega_0)t-i (\vk -\vk_0)\cdot\vr_p(t)}\, h(t) \int \mathrm d^3u\ \frac{e^{-i (\vk-\vk_0) \cdot\vu}}{u}\\
    &=-G M \sqrt \rho \int \mathrm d t\ e^{i (\omega-\omega_0)t-i (\vk -\vk_0)\cdot\vr_p(t)}\, h(t)\, \frac{4 \pi}{|\vk-\vk_0|^2}\nonumber\;,
\end{align}
where $\vu=\vr-\vr_p(t)$ and, in the last equality, the Fourier transform of the Coulomb potential is used. Together with the Fourier transform of the retarded Green's function, this leads to
\begin{align}
    \delta \psi (\vr,t ) &=m_a \intk\intw e^{-i \omega t+i \vk\cdot\vr}\, \tilde G(\vk, \omega)\, \Tilde{(V\psi_0)}(\vk ,\omega) \\
    &=4 \pi m_a M G \lim_{\epsilon\rightarrow 0^+} \int \mathrm d t'\ h(t')\,\sqrt \rho\, e^{-i \omega_0 t'+i \vk_0\cdot\vr_p(t')}
     \intk\intw\ \frac{e^{-i \omega(t-t')+i\vk\cdot(\vr-\vr_p(t'))}}{\hbar (\omega+i \epsilon) -\frac{\hbar^2 k^2}{2 m_a}}\,\frac{1}{|\vk-\vk_0|^2}\nonumber\\
    &=4 \pi m_a M G\lim_{\epsilon\rightarrow 0^+}\int \mathrm d t'\ h(t')\, \psi_0(\vr_p(t'),t') \intk\intw\ \frac{e^{-i \omega\tau+i\vk\cdot\vu(t') }}{\hbar (\omega+i \epsilon) -\frac{\hbar^2 k^2}{2 m_a}}\,\frac{1}{|\vk-\vk_0|^2}\nonumber\;.
\end{align}
Once again the Cauchy integration formula is used to solve the $\omega$-integral. There is only one pole at $\omega=\frac{\hbar}{2m_a}k^2-i \epsilon$, which is in the lower half of the complex plane. For the integral to be non-zero,
a contour through the lower half plane has to be chosen (similar to C2 in the left panel of Fig.~\ref{fig:polesSll}), and the arc through the complex plane only gives vanishing contribution when $\tau>0$. This ensures causality.
The solution to the $\omega$-integral is given by
\begin{equation}
    \intw\ \frac{e^{-i \omega\tau}}{\hbar (\omega +i \epsilon) -\frac{\hbar^2 k^2}{2 m_a}}=-\frac{i}{\hbar}H(\tau)e^{-i \tau (\frac{\hbar k^2}{2m_a}-i \epsilon)}\;,
\end{equation}
at which point we can safely take the limit $\epsilon\rightarrow0$. Inserting this result into the expression for $\delta\psi$, the latter can be further simplified to
\begin{align}
\delta \psi (\vr,t ) &=i \frac{4 \pi G M m_a }{\hbar}\int \mathrm d t'\ h(t')\,H(\tau)\, \psi_0(\vr_p(t'),t') \intk\ e^{i\vk\cdot\vu(t')-i \tau \beta k^2}\frac{1}{|\vk -\vk_0|^2}\\
&=i \frac{4 \pi G M m_a }{\hbar}\int \mathrm d t'\ h(t')\,H(\tau)\, \psi_0(\vr_p(t'),t') \int_{\mathbf k'}\ e^{i(\vk'+\vk_0)\cdot\vu(t')-i \tau \beta (\vk'+\vk_0)^2}\frac{1}{ k'^2}\nonumber\\
&=i \frac{4 \pi G M m_a}{\hbar}\int \mathrm d t'\ h(t')\,H(\tau)\, \psi_0(\vr_p(t'),t')\,e^{i \vk_0\cdot\vu(t')-i \tau \beta k_0^2}\int_{\mathbf k'}\ e^{i \vk'\cdot(\vu(t')-2\tau\beta \vk_0)-i \tau \beta k'^2}\frac{1}{k'^2}
\nonumber
\end{align}
upon substituting $\vk'=\vk-\vk_0$ and the diffusion coefficient $\beta=\hbar/2m_a$. The first part is independent of $\vk'$ and, with help of the dispersion relation Eq.~(\ref{eq:disp}), simplifies to $\psi_0(\vec r,t)$ which is independent of the remaining integration over $\vk'$. Performing the latter leads to
\begin{align}
    \intk\ e^{i \vk'\cdot(\vu(t')-2\tau\beta \vk_0)-i \tau \beta k'^2}\frac{1}{k'^2}
    &=\frac{1}{2\pi^2}\int \mathrm d k'\ j_0\big(k'|\vu(t')-2\tau\beta\vk_0|\big)\, e^{-i \tau \beta k'^2}\\
    &=\frac{1}{2\pi^2}\, \frac{\pi}{2}(1+i)\,\frac{\left[\textit{S}\!\left(\frac{|\vu(t')-2\tau\beta \vk_0|}{\sqrt{2\pi \beta \tau}}\right)-i \textit{C}\!\left(\frac{|\vu(t')-2\tau\beta \vk_0|}{\sqrt{2\pi \beta \tau}}\right)\right]}{|\vu(t')-2\tau\beta \vk_0|}\nonumber\;,
\end{align}
where $\textit{C}(z)$ and $\textit{S}(z)$ are the Fresnel integrals. Using their connection with the Error function,
 \begin{align}
     \textit{S}(z)&=\frac{(1+i)}{4}\left[\mathrm{erf}\!\left(\frac{1+i}{2}\sqrt{\pi}z\right)-i\ \mathrm{erf}\!\left(\frac{1-i}{2}\sqrt{\pi}z\right)\right]\\
     \textit{C}(z)&=\frac{(1-i)}{4}\left[\mathrm{erf}\!\left(\frac{1+i}{2}\sqrt{\pi}z\right)+i\ \mathrm{erf}\!\left(\frac{1-i}{2}\sqrt{\pi}z\right)\right] \nonumber \;,
 \end{align}
 we eventually arrive at
 \begin{equation}
    \delta \psi(\vr,t)=i \frac{G M m_a}{\hbar}\,\psi_0(\vr,t) \int \mathrm d t'\ h(t')\,H(\tau)\,\frac{\mathrm{erf}\!\left(\frac{1-i}{2} \frac{|\vu(t')-2\tau\beta \vk_0|}{\sqrt{2 \beta \tau}} \right)}{|\vu(t')-2\tau\beta \vk_0|}
\end{equation}
which gives Eq.~(\ref{eq:delPSi}) after substituting the dimensionless variables.

\section{Scattering amplitudes in the steady-state regime}
\label{sec:detStdy}

In this Appendix, we provide details of the calculation of the scattering amplitudes $\slls$ in the steady-state regime, beginning with Eq.~(\ref{eq:sllStart}). 

First, the spherical Bessel functions is split into a sum of Hankel functions $\bessJ{l}{z}=\frac{1}{2}(\hone{l}{z}+\htwo{l}{z})$, and the integral acquires a factor of $1/2$ while its limits are extended to $-\infty$ to $\infty$. Since $\hone{l}{z}\propto e^{i z}$ and $\htwo{l}{z}\propto e^{-i z}$, each of the terms resulting from the product of the two spherical Bessel functions is proportional to a complex exponential with a positive, zero or negative phase which determines the contour to be chosen. 

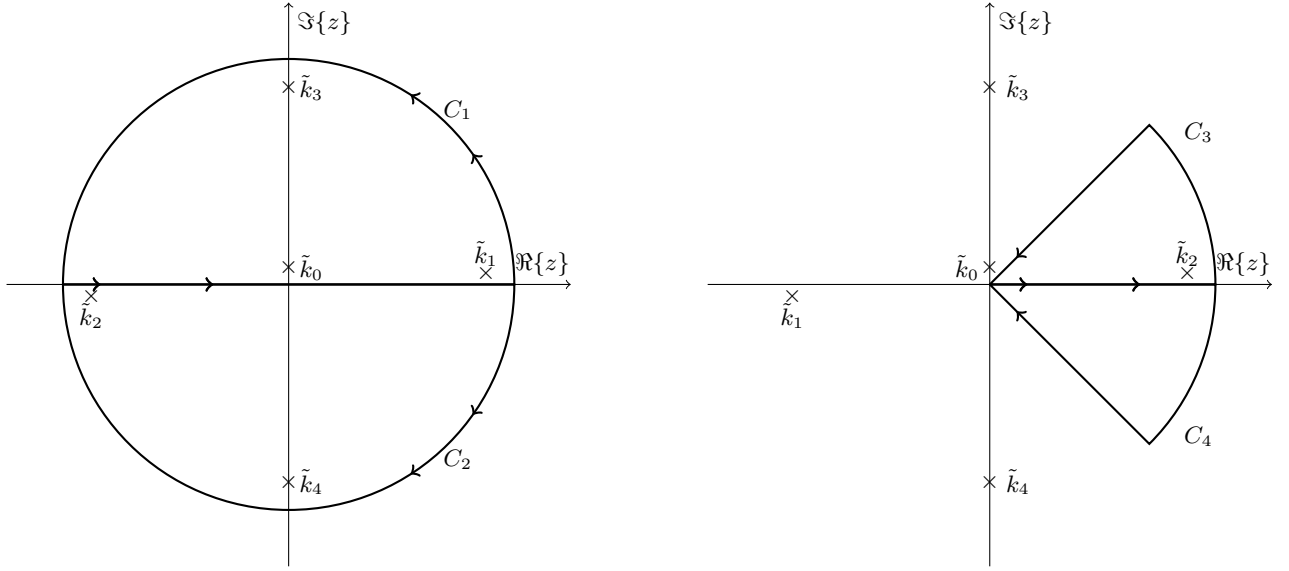
\begin{figure}
    \begin{minipage}{.48\textwidth}
        \begin{tikzpicture}[decoration={markings,
mark=at position 0.5cm with {\arrow[line width=1pt]{>}},
mark=at position 2cm with {\arrow[line width=1pt]{>}},
mark=at position 7.85cm with {\arrow[line width=1pt]{>}},
mark=at position 9cm with {\arrow[line width=1pt]{>}}
},scale=0.75
]
\draw [->] (-5,0) -- (5,0) node [above left,xshift=0.1cm]  {\small$\Re\{z\}$};
\draw [->] (0,-5) -- (0,5) node [below right] {\small$\Im\{z\}$};

\node at (0,0.3){$\times$};
\node at (.4,.3){$\tk_0$};
\node at (-3.5,-0.2) {$\times$};
\node at (-3.5,-0.55) {$\tk_2$};
\node at (3.5,0.2) {$\times$};
\node at (3.5,0.55) {$\tk_1$};
\node at (0,3.5){$\times$};
\node at (.4,3.5){$\tk_3$};
\node at (0,-3.5){$\times$};
\node at (0.4,-3.5){$\tk_4$};

\node at (3,3.1){$C_1$};
\node at (3,-3.1){$C_2$};

\path[draw,line width=0.8pt,postaction=decorate] (-4,0)-- (4,0) arc (0:180:4) ;
\path[draw,line width=0.8pt,postaction=decorate] (-4,0)-- (4,0) arc (0:-180:4) ;

\end{tikzpicture}
    \end{minipage}
    \hfill
    \begin{minipage}{.48\textwidth}
        \begin{tikzpicture}[decoration={markings,
mark=at position 0.5cm with {\arrow[line width=1pt]{>}},
mark=at position 2cm with {\arrow[line width=1pt]{>}},
mark=at position 7.85cm with {\arrow[line width=1pt]{>}},
mark=at position 9cm with {\arrow[line width=1pt]{>}}
}
,scale=0.75]
\draw [->] (-5,0) -- (5,0) node [above left,xshift=0.1cm]  {\small$\Re\{z\}$};
\draw [->] (0,-5) -- (0,5) node [below right] {\small$\Im\{z\}$};

\node at(0,.3){$\times$};
\node at (-.4,.3){$\tk_0$};
\node at (-3.5,-0.2) {$\times$};
\node at (-3.5,-0.55) {$\tk_{1}$};
\node at (3.5,0.2) {$\times$};
\node at (3.5,0.55) {$\tk_{2}$};
\node at (0,3.5){$\times$};
\node at (.5,3.5){$\tk_{3}$};
\node at (0,-3.5){$\times$};
\node at (0.5,-3.5){$\tk_{4}$};

\node at (3.7,2.7){$C_3$};
\node at (3.7,-2.7){$C_4$};

\path[draw,line width=0.8pt,postaction=decorate] (0,0)-- (4,0) arc (0:45:4)--(0,0);
\path[draw,line width=0.8pt,postaction=decorate] (0,0)-- (4,0) arc (0:-45:4)--(0,0);

\end{tikzpicture}
    \end{minipage}
    \caption{{\it Left panel}: poles and semi-circular contours $C_1$ and $C_2$ used for the calculation of the steady-state amplitude Eq.~(\ref{eq:sllStart}). The poles are located at $\tilde k_0=0$, $\tilde k_{1,2}=\pm \sqrt{m\rv+i \epsilon}$ and $\tilde k_{3,4}=\pm i \sqrt{m\rv+i \epsilon}$. We choose to shift the $\tilde k=\tilde k_0$ pole into the upper half plane. The contour $C_1$ ($C_2$) extents from $-\infty$ to $\infty$ on the real axis and is completed in the upper (lower) half plane. {\it Right panel}: The contours $C_3$ and $C_4$ used for the calculation of the transient amplitude Eq.~(\ref{eq:slltOrig}). For $C_3$ ($C_4$), the circular arc subtends an angle $\pi/4$ and lies in the upper (lower) half plane. The two contours are closed by a diagonal line at an angle $\vartheta=\pm\frac{\pi}{4}$ which can be parameterized by $k=(1\pm i)\chi$ with $\chi$ in the range $[0,\infty[$.}
    \label{fig:polesSll}
\end{figure}

Poles are found at $\tk_0=0$, $\tk_{1/2}=\pm \sqrt{m\rv +i \epsilon}$ and $\tk_{3/4}=\pm i \sqrt{m\rv+i \epsilon}$ as indicated in the left panel of Fig.~\ref{fig:polesSll}.  $\tk_0$, $\tk_1$ and $\tk_3$ are enclosed in the contour $C_1$ relevant for the terms with a positive phase. It is trivial to show that the contour integral over the semi-circle vanishes. Similarly, the contour $C_2$, which contains the poles at $k_2$ and $k_4$, must be selected for the terms with a negative phase. The choice of $C_2$ is also more convenient when the phase is zero. 

The residues are independent of the exact combination of $\hone{l}{z}$ and $\htwo{l}{z}$. Therefore, we introduce below the generic notation $(x)$ or $(y)$ to label the Hankel functions. Both $(x)$ and $(y)$ can be either $(1)$ or $(2)$. Furthermore. we use the shorthand notation $(\bar x)$ to indicate that a $\hone{l}{z}$ was transformed into a $\htwo{l}{z}$ or conversely through the relation $\bh{1}{l}{-z}=(-1)^{l}\bh{2}{l}{z}$. Taking the limit $\epsilon\to 0^+$, we have:
\begin{align}
    \res{\green{x}{y}}{\tk_0}&=\frac{i}{m^2}\\
    \res{\green{x}{y}}{\tk_1}&=\resResult{x}{y}{\rwb} \nonumber \\
    \res{\green{x}{y}}{\tk_2}&=-\resResult{\bar x}{\bar y}{\rwb} \nonumber\\
    \res{\green{x}{y}}{\tk_3}&=-\resResult{x}{y}{i\rwb}\nonumber\\
    \res{\green{x}{y}}{\tk_4}&=\resResult{\bar{x}}{\bar{y}}{i \rwb} \nonumber\;.
\end{align}
All of these contributions combine to give 
\begin{align}
    \slls &=\frac{i\pi}{4}\frac{\rv}{4 m}\bigg[\frac{4 i}{m\rv}+2\hone{l}{\rwb}\,\hone{l-1}{\rwb}
        +\hone{l}{\rwb}\,\htwo{l-1}{\rwb}\\ 
        &\qquad +\htwo{l}{\rwb}\,\hone{l-1}{\rwb}
        -(\rwb \Leftrightarrow i \rwb)\bigg] \nonumber \;,
\end{align}
where $(\rwb \Leftrightarrow i \rwb)$ indicates that all the terms involving Hankel functions are repeated but with their argument replaced by $i\rwb$. Using $\hone{l}{x}=\bessJ{l}{x}+i \bessY{l}{x}$, $\htwo{l}{x}=\bessJ{l}{x}-i \bessY{l}{x}$, as well as the Wronskian relation $\bessJ{l}{x}\bessY{l-1}{x}-\bessJ{l-1}{x}\bessY{l}{x}=x^{-2}$, the sum of terms with argument $q=\rwb$ can be simplified to 
\begin{align*}
     2\hone{l}{q}\,\hone{l-1}{q}&+\hone{l}{q}\,\htwo{l-1}{q} +\htwo{l}{q}\,\hone{l-1}{q}+\frac{2 i}{q^2} \\
    &=4\bessJ{l}{q}\,\bessJ{l-1}{q}+4i\bessJ{l}{q}\,\bessY{l-1}{q}- 2i (\bessJ{l}{q}\,\bessY{l-1}{q}-\bessJ{l-1}{q}\,\bessY{l}{q})+\frac{2 i}{q^2}\\
    &=\Big. 4\bessJ{l}{q}\,(\bessJ{l-1}{q}+i \bessY{l-1}{q})\\
    &=\Big. 4\bessJ{l}{q}\,\hone{l-1}{q}\;,
\end{align*}
and likewise for the terms with argument $q=i \rwb$. These simplifications yield
\begin{equation}
    \slls=\frac{i \pi\rv}{4 m}\left[\bessJ{l}{\rwb}\,\hone{l-1}{\rwb}-\bessJ{l}{i \rwb}\,\hone{l-1}{i \rwb}\right] \;.
\end{equation}
For the final step of the calculation, the identities $\bessJ{l}{i x}=i^l \textit{i}^{(1)}_l(x)$ and $\hone{l}{i x}=-\frac{2}{\pi}i^l\textit{k}_{l}(x)$ are exploited and lead to Eq.~(\ref{eq:slls}),
\begin{equation*}
    \slls
    =\frac{i \pi\rv}{4m}\left[\bessJ{l}{\rwb}\,\hone{l-1}{\rwb}+\frac{1}{\sqrt{m\rv}}\textit{I}_{l+1/2}\left(\rwb\right)\,\textit{K}_{l-1/2}\left(\rwb\right)\right]\;,
\end{equation*}
upon substituting $\textit{I}^{(1)}_l(x)=\sqrt{\frac{\pi}{2 x}}\textit{I}_{l+1/2}$ and $\textit{K}_{l}(x)=\sqrt{\frac{\pi}{2 x}}\textit{K}_{l+1/2}(x)$ for numerical convenience.

\section{Scattering amplitudes for the finite time perturbation}
\label{sec:detFin}

For the finite time perturbation, we insert Eq.~(\ref{eq:transStart}) into Eq.~(\ref{eq:scat}) and obtain
\begin{equation}
      \sllf=\lim_{\epsilon\rightarrow0^+}\intw \frac{e^{i (m-\tilde\omega)\ttau}}{i(m-\tilde\omega-i \eta)}\int_0^\infty \mathrm{d}\tk\  \tk \frac{\bessJ{l}{\tk}\bessJ{l-1}{\tk}}{\tk^4/\rv^2-(\omega+i\epsilon)^2}\;,
\end{equation}
where it is understood that $\eta>0$.
In a first step, the integral over $\tilde \omega$ is solved with the Residue theorem. We choose the contour $C_2$ as in the left panel of Fig.~\ref{fig:polesSll}, with a semicircular arc in the lower half of the complex plane. There are three poles at $\tilde\omega_0=m-i \eta$, $\tilde \omega_1=\tk^2/\rv-i\epsilon$ and $\tilde \omega_2=-\tk^2/\rv-i \epsilon$. They are all located in the lower half plane and, therefore, all contribute as Residues. Taking $\eta\to 0$ yields
\begin{align}
    \sllf &=-\lim_{\epsilon\rightarrow 0^+}\int_0^\infty \dtk\ \tk\, \bessJ{l}{ \tk}\,\bessJ{l-1}{ \tk}\, e^{i m \tilde t}\, \Bigg[\frac{e^{i m\tilde t}}{(m+i\epsilon)^2-\tk^4/\rv^2}\\ &\qquad
    +\frac{\rv}{2 \tk^2}\left( \frac{e^{-i(\tk^2/\rv-i \epsilon)\tilde t}}{\tk^2/\rv-m-i\epsilon}-\frac{e^{-i(-\tk^2/\rv-i \epsilon)\tilde t}}{-\tk^2/\rv-m-i\epsilon}\right)\Bigg]\nonumber\;.
\end{align}
Upon taking the limit $\eta\to 0$, the first term in the square brackets can be identified as the steady-state amplitude Eq.~(\ref{eq:sllStart}). It is discussed in Sec.~\S\ref{sec:detStdy} and we shall thus ignore it here. The second term defines the transient amplitude $\sllt$:
\begin{equation}
\label{eq:slltOrig}
     \sllt\equiv-\frac{\rv}{2} e^{i m\tilde t}\int_0^\infty \frac{\dtk}{\tk} \bessJ{l}{\tk}\,\bessJ{l-1}{\tk}\, \Bigg(\frac{e^{-i(\tk^2/\rv-i \epsilon)\tilde t}}{\tk^2/\rv-m-i\epsilon}+\frac{e^{i(\tk^2/\rv+i \epsilon)\tilde t}}{\tk^2/\rv+m +i\epsilon}\Bigg)\;.
\end{equation}
To compute this amplitude, we use the contours $C_3$ and $C_4$ in the upper and lower half plane, respectively, as indicated in the right panel of Fig.~\ref{fig:polesSll}. The piece proportional to $e^{i\tk^2\tilde t/\rv}$ is evaluated with $C_3$. However, the relevant poles $\tilde k_3$ and $\tilde k_4$ are not enclosed by $C_3$. Cauchy's integral formula thus implies that its line integral is zero. Since the integral over the circular arc vanishes when its radius tends to infinity, this also implies that the integral over the positive real axis $\tilde k\in[0,+\infty[$ is minus the line integral over the diagonal parameterized by $\tilde k = (1+i)\chi$, $\chi\in[0,+\infty[$. The same reasoning applies to the contribution proportional to $e^{-i\tk^2\tilde t/\rv}$, the contour $C_4$ and the relevant poles $\tilde k_1$ and $\tilde k_2$. Adding up the two contributions eventually gives Eq.~(\ref{eq:sllt})~:
\begin{equation}
    \sllt=-\frac{\rv}{2}e^{i m\tilde t}\int_0^\infty\frac{\mathrm{d}\tilde\chi}{\tilde\chi}\Big[ \bessJ{l}{(1+i)\tilde\chi}\bessJ{l-1}{(1+i)\tilde\chi}-\bessJ{l}{(1-i)\tilde \chi}\bessJ{l-1}{(1-i)\tilde\chi}\Big]\frac{e^{-2\tilde t \tilde\chi^2/\rv^2}}{2 i\tilde\chi^2/\rv+m+i\epsilon}
\end{equation}
This expression makes clear that $\sllt$ must tend to zero in the limit $\tilde t\to\infty$ so long as $m\ne 0$.

\section{Analytic expression for $\sllb{a}{b}(m,\rv)$ with $(l,m)=(1,0)$}
\label{sec:sllbl1}

They can be obtained with a software such as {\small MATHEMATICA}~\cite{Mathematica}. For $q_1>q_2$, we have
\begin{align}
    \sllb{a}{b}(0,\rv)&=\frac{\rv^2}{240 \tkmin^5 q_1^2 q_2}\bigg\{\pi  \tkmin^5 q_2 \Big(-15 q_1^4-10 q_1^2 q_2^2+q_2^4\Big)\\
    &\quad +
    \tkmin^5 \left[(3 q_2-4) (1-2 q_2)^4 \text{Si}(\tkmin-2 \tkmin q_2)+(4-5 q_2) \text{Si}(\tkmin)\right] \nonumber \\
    &\quad +
    \tkmin q_1 \cos (\tkmin q_1) \left[2 \left(\tkmin^2 \left(2 q_1^2+q_2^2\right)-12\right) \sin (\tkmin q_2)+\tkmin q_2 \left(\tkmin^2 \left(11 q_1^2+q_2^2\right)-6\right) \cos (\tkmin q_2)\right] \nonumber \\
    &\quad +
    2 \sin (\tkmin q_1) \bigg[\tkmin q_2 \left(\tkmin^2 \left(7 q_1^2-q_2^2\right)+6\right) \cos (\tkmin q_2) \nonumber \\
    &\qquad+
    \left(\tkmin^4 \left(-4 q_1^4-9 q_1^2 q_2^2+q_2^4\right)+\tkmin^2 \left(8 q_1^2-2 q_2^2\right)+24\right) \sin (\tkmin q_2)\bigg]\bigg\} \nonumber
\end{align}
whereas, for $q_2>q_1$, we obtain
\begin{align}
\sllb{a}{b}(0,\rv) &=\frac{\rv^2}{240 \tkmin^5 q_1^2 q_2}\bigg\{ 4 \pi  \tkmin^5 (q_2-1)^3 \Big(6 q_2^2-2 q_2+1\Big)\\
&\quad+
\tkmin^5 \left[(3 q_2-4) (1-2 q_2)^4 \text{Si}(\tkmin-2 \tkmin q_2)+(4-5 q_2) \text{Si}(\tkmin)\right] \nonumber \\
&\quad+
2\tkmin \cos (\tkmin q_1) \left[4 \left(-3 \tkmin^2 q_2+\tkmin^2+6 q_2-6\right) \sin (\tkmin q_2)+\tkmin \left(11 \tkmin^2-6\right) q_2 \cos (\tkmin q_2)\right]\nonumber \\
&\quad+
2\sin (\tkmin q_1) \left[4 \left(\tkmin^2 \left(\tkmin^2 (4 q_2-1)-4 q_2+2\right)+6\right) \sin (\tkmin q_2)+\tkmin \left(7 \tkmin^2+6\right) q_2 \cos (\tkmin q_2)\right] \nonumber \\
&\quad+2\tkmin^2 q_2^2 \left[\left(\tkmin^2 (2 (17-6 q_2) q_2-33)+6\right) \cos (\tkmin-2 \tkmin q_2)+2 \tkmin (3 q_2-7) \sin (\tkmin-2 \tkmin q_2)\right]\bigg\} \nonumber
\end{align}

\end{document}